\documentclass[10pt,journal,compsoc]{IEEEtran}
\usepackage[boxed,ruled,vlined,linesnumbered]{algorithm2e}
\usepackage{multirow}
\usepackage{rotating}
\usepackage{hhline}
\usepackage{url}
\usepackage{framed}
\usepackage{hyperref}
\usepackage{longtable}
\newcommand*\wrapletters[1]{\wr@pletters#1\@nil}
\def\wr@pletters#1#2\@nil{#1\allowbreak\if&#2&\else\wr@pletters#2\@nil\fi}
\usepackage{tikz}
\usetikzlibrary{automata,matrix,shapes,arrows,positioning,chains,calc}
\usetikzlibrary{snakes}
\usetikzlibrary{arrows,scopes}
\usetikzlibrary{positioning,chains,fit,shapes,calc}
\usepackage{caption}
\usepackage{subcaption}
\usepackage{amsmath}
\SetKwHangingKw{Local}{Local}
\SetKwHangingKw{Global}{Global}
\SetKwHangingKw{Constant}{Constant}
\SetKwHangingKw{Input}{Input}
\SetKwHangingKw{Alias}{Alias}
\SetKwHangingKw{Output}{Output}
\SetKwHangingKw{SideEffect}{Side effects}
\SetKwHangingKw{InternalEvent}{Internal event}
\SetKwHangingKw{External}{External}
\SetKwHangingKw{ExternalEvent}{External event}
\SetKwHangingKw{Precondition}{Precondition :}
\SetKwHangingKw{Postcondition}{Postcondition :}
\SetKwHangingKw{Upon}{Upon}
\SetKwHangingKw{DoForever}{Do Forever}
\SetKwHangingKw{PVab}{Persistent variables:}

\usepackage{theorem}
\usepackage{wrapfig}
\usepackage{amssymb}
\usepackage{xspace}

\newcommand{\qed}{\hfill $\blacksquare$}
\newcommand{\qedClaim}{\hfill \ensuremath{\Box}}

\newcommand{\B}{\vspace*{-\smallskipamount}}
\newcommand{\BB}{\vspace*{-\medskipamount}}
\newcommand{\BBB}{\vspace*{-\bigskipamount}}

\newcommand{\remove}[1]{}

\usepackage{csquotes}
\usepackage{enumitem}
\usepackage{tablefootnote}
\usepackage{fnpct}
\usepackage[mathlines,switch]{lineno}

\ifCLASSOPTIONcompsoc
  \usepackage[nocompress]{cite}
\else
  \usepackage{cite}
\fi
\ifCLASSINFOpdf
\else
\fi

\IEEEoverridecommandlockouts

\begin{document}
\title{A Survey on Geographically Distributed Big-Data Processing using MapReduce}

\author{Shlomi Dolev,~\IEEEmembership{Senior Member,~IEEE,}
        Patricia Florissi, 
        Ehud Gudes,~\IEEEmembership{Member, IEEE Computer Society,}
        Shantanu Sharma,~\IEEEmembership{Member,~IEEE,}
        and~Ido Singer
\IEEEcompsocitemizethanks{\IEEEcompsocthanksitem S. Dolev is with the Department of Computer Science, Ben-Gurion University of the Negev, Israel. E-mail: \texttt{dolev@cs.bgu.ac.il}
\IEEEcompsocthanksitem P. Florissi is with Dell EMC, USA. E-mail: \texttt{patricia.florissi@dell.com}
\IEEEcompsocthanksitem E. Gudes is with the Department of Computer Science, Ben-Gurion University of the Negev, Israel. E-mail: \texttt{ehud@cs.bgu.ac.il}
\IEEEcompsocthanksitem S. harma is with the Department of Computer Science, University of California, Irvine, USA. E-mail: \texttt{shantanu.sharma@uci.edu}
\IEEEcompsocthanksitem I. Singer is with Dell EMC, Israel. E-mail: \texttt{ido.singer@dell.com}}
\thanks{Manuscript received 31 Oct. 2016; accepted 30 June 2017. DOI: 10.1109/TBDATA.2017.2723473. \copyright 2017 IEEE. Personal use of this material is permitted. Permission from IEEE must be obtained for all other uses, including reprinting/republishing this material for advertising or promotional purposes, collecting new collected works for resale or redistribution to servers or lists, or reuse of any copyrighted component of this work in other works. 

The final version of this paper may differ from this accepted version.}
}


\markboth{IEEE Transactions on Big Data; Accepted June 2017.}%
{Dolev \MakeLowercase{\textit{et al.}}: A Survey on Geographically Distributed Big-Data Processing using MapReduce}
\IEEEtitleabstractindextext{
\begin{abstract}
Hadoop and Spark are widely used distributed processing frameworks for large-scale data processing in an efficient and fault-tolerant manner on private or public clouds. These big-data processing systems are extensively used by many industries, \textit{e}.\textit{g}., Google, Facebook, and Amazon, for solving a large class of problems, \textit{e}.\textit{g}., search, clustering, log analysis, different types of join operations, matrix multiplication, pattern matching, and social network analysis. However, all these popular systems have a major drawback in terms of \emph{locally distributed} computations, which prevent them in implementing geographically distributed data processing. The increasing amount of geographically distributed massive data is pushing industries and academia to rethink the current big-data processing systems. The novel frameworks, which will be beyond state-of-the-art architectures and technologies involved in the current system, are expected to process geographically distributed data at their locations without moving entire \emph{raw datasets} to a single location. In this paper, we investigate and discuss challenges and requirements in designing geographically distributed data processing frameworks and protocols. We classify and study batch processing (MapReduce-based systems), stream processing (Spark-based systems), and SQL-style processing geo-distributed frameworks, models, and algorithms with their overhead issues.
\end{abstract}

}
\maketitle


\IEEEdisplaynontitleabstractindextext
\IEEEpeerreviewmaketitle
\ifCLASSOPTIONcompsoc
\IEEEraisesectionheading{\section{Introduction}\label{sec:introduction}}
\else

\section{Introduction}
\label{sec:introduction}
\fi
\IEEEPARstart{C}urrently, several cloud computing platforms, \textit{e}.\textit{g}., Amazon Web Services, Google App Engine, IBM's Blue Cloud, and Microsoft Azure, provide an easy \textit{locally distributed}, scalable, and on-demand big-data processing. However, these platforms do not regard geo(graphically) data locality, \textit{i}.\textit{e}., geo-distributed data~\cite{DBLP:conf/ic2e/JonathanCW16}, and hence, necessitate data movement to a single location before the computation.

In contrast, in the present time, data is generated geo-distributively at a much higher speed as compared to the existing data transfer speed~\cite{DBLP:conf/cidr/VulimiriCGKV15,DBLP:journals/cacm/ReedD15}; for example, data from modern satellites~\cite{DBLP:journals/pc/HawickCJ03}. There are two common reasons for having geo-distributed data, as follows: (\textit{i}) Many organizations operate in different countries and hold datacenters (DCs) across the globe. Moreover, the data can be distributed across different systems and locations even in the same country, for instance, branches of a bank in the same country. (\textit{ii}) Organizations may prefer to use multiple public and/or private clouds to increase reliability, security, and processing~\cite{DBLP:journals/pvldb/KloudasRPM15,url2,url3}. In addition, there are several applications and computations that process and analyze a huge amount of massively geo-distributed data to provide the final output. For example, a bioinformatic application that analyzes existing genomes in different labs and countries to track the sources of a potential epidemic. The following are few examples of applications that process geo-distributed datasets: climate science~\cite{DBLP:conf/ccgrid/TudoranCWBA14,DBLP:conf/ipps/TudoranAB13}, data generated by multinational companies~\cite{akella-15,DBLP:conf/ccgrid/TudoranCWBA14,DBLP:conf/IEEEcloud/ZhangLLZSMGA14}, sensor networks~\cite{DBLP:conf/ipps/TudoranAB13,DBLP:conf/hotos/RabkinASPF13}, stock exchanges~\cite{DBLP:conf/ipps/TudoranAB13}, web crawling~\cite{cardosa2011exploring,heintz2015end}, social networking applications~\cite{cardosa2011exploring,heintz2015end}, biological data processing~\cite{DBLP:conf/ccgrid/TudoranCWBA14,DBLP:conf/hotos/RabkinASPF13,CPE15} such as DNA sequencing and human microbiome investigations, protein structure prediction, and molecular simulations, stream analysis~\cite{DBLP:conf/ipps/TudoranAB13}, video feeds from distributed cameras, log files from distributed servers~\cite{DBLP:conf/hotos/RabkinASPF13}, geographical information systems (GIS)~\cite{DBLP:journals/pc/HawickCJ03}, and scientific applications~\cite{cardosa2011exploring,DBLP:conf/ccgrid/TudoranCWBA14,g-hadoop,DBLP:conf/ipps/TudoranAB13}.

It should be noted down here that all the above-mentioned applications generate a high volume of \emph{raw data} across the globe; however, most analysis tasks require only a small amount of the original raw data for producing the final outputs or summaries~\cite{DBLP:conf/hotos/RabkinASPF13}.

\noindent\textbf{Geo-distributed big-data processing vs. the state-of-the-art big-data processing frameworks.} Geo-distributed databases and systems have been in existence for a long time~\cite{DBLP:journals/csur/ShethL90}. However, these systems are not highly fault-tolerant, scalable, flexible, good enough for massively parallel processing, simple to program, able to process a large-scale (and/or real-time) data, and fast in answering a query.

On a positive side, several \textit{big-data processing programming models and frameworks} such as MapReduce~\cite{DBLP:conf/osdi/DeanG04}, Hadoop~\cite{apache_hadoop}, Spark~\cite{spark}, Dryad~\cite{DBLP:conf/eurosys/IsardBYBF07}, Pregel~\cite{DBLP:conf/sigmod/MalewiczABDHLC10}, and Giraph~\cite{Giraph} have been designed to overcome the disadvantages (\textit{e}.\textit{g}., fault-tolerance, unstructured/massive data processing, or slow processing time) of parallel computing, distributed databases, and cluster computing. Thus,
this survey paper focuses on the MapReduce, Hadoop, and Spark based systems. On a negative side, these frameworks do not regard geo-distributed data locations, and hence, they follow a trivial solution for geo-distributed data processing: copy all \emph{raw data} to one location before executing a \textit{locally distributed} computation.

The trivial solution has a bottleneck in terms of data transfer, since it is not always possible to copy the whole \emph{raw data} from different locations to a single location due to security, privacy, legal restrictions, cost, and network utilization. Moreover, if the output of the computation at each site is smaller than the input data, it is completely undesirable to move the raw input data to a single location~\cite{DBLP:journals/tc/JayalathSE14,cardosa2011exploring,DBLP:journals/sigmetrics/GadreRP11}. In a widely distributed environment with network heterogeneity, Hadoop does not work well because of heavy dependency between MapReduce phases, highly coupled data placement, and task execution~\cite{DBLP:conf/ic2e/HeintzWCW13}. In addition, HDFS Federation~\cite{hdfs-f} cannot support geo-distributed data processing, because DataNodes at a location are not allowed to register themselves at a NameNode of another location, which is governed by another organization/country. Thus, the current systems cannot process data at multiple-clusters.

It is also important to mention that the network bandwidth is also a crucial factor in geo-distributed data movement. For example, the demand for bandwidth increased from 60Tbps to 290Tbps between the years 2011 and 2015 while the network capacity growth was not proportional. In the year 2015, the network capacity growth was only 40\%, which was the lowest during the years 2011 and 2014~\cite{network}.

\begin{figure}[!h]
\centering
\B
\includegraphics[scale=0.28]{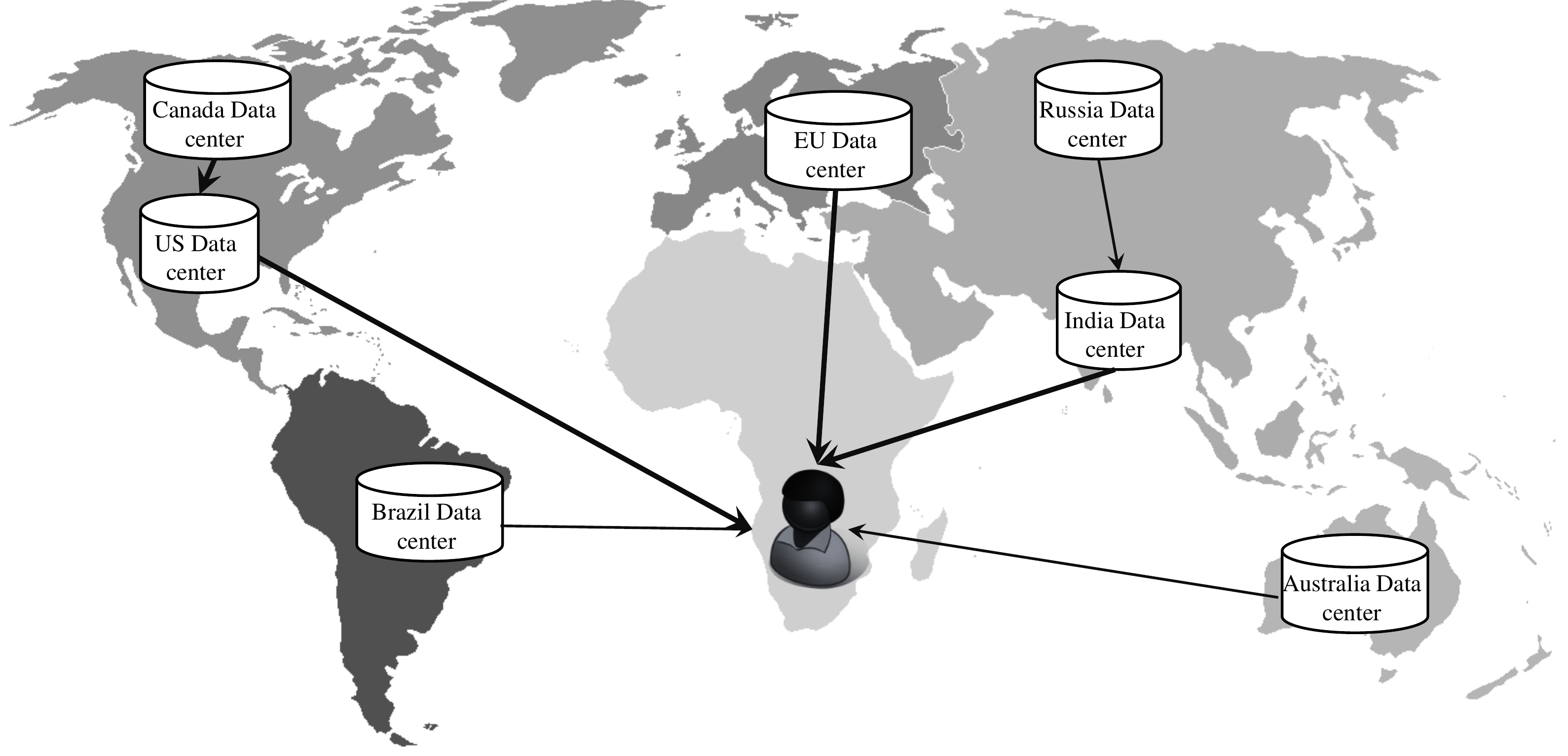}
\BBB
\caption{A scenario for geographically distributed data processing.}
\label{fig:distributed_processing}
\BB
\end{figure}

Fig.~\ref{fig:distributed_processing} shows an abstract view of desirable geo-distributed data processing, where different locations hold data and a local computation is executed on the site. Each site executes an assigned computation locally distributed and transfers (partial) outputs to the closest site or the user site. Eventually, all the (partial) outputs are collected at a single site (or the user site) that executes another job to obtain the final outputs. Different sites are connected with different speeds (the bandwidth consideration in the context of geo-distributed data processing is given in~\cite{akella-15}). The thick lines show high bandwidth networks, and the thinner lines are lower bandwidth networks.

Currently, several researchers are focusing on the following important questions: what can be done to process geo-distributed big-data using Hadoop, Spark, and similar frameworks, and how? Can we process data at their local sites and send only the outputs to a single location for producing the final output? The on-site processing solution requires us to rethink, redesign, and revisualize the current implementations of Hadoop, Spark, and similar frameworks. In this work, we will review several models, frameworks, and resource allocation algorithms for geo-distributed big-data processing that try to solve the above-mentioned problems. In a nutshell, geo-distributed big-data processing frameworks have the following properties:
\begin{itemize}[nolistsep,noitemsep,leftmargin=0.14in]
  \item \textit{Ubiquitous computing}: The new system should regard different data locations, and it should process data at different locations, transparent to users. In other words, new geo-distributed systems will execute a geo-computation like a locally distributed computation on geo-locations and support any type of big-data processing frameworks, languages, and storage media at different locations~\cite{g-hadoop,DBLP:journals/tc/JayalathSE14,akella-15,hog}.

  \item \textit{Data transfer among multiple DCs}: The new system should allow moving only the \emph{desired data}, which eventually participate in the final output,\footnote{We are not explaining the method of finding only desired data before the computation starts. Interested readers may refer to~\cite{metamr_sss-15}.} in a secure and privacy-preserving manner among DCs, thereby reducing the need for high bandwidth~\cite{akella-15,hmr,hog}.

  \item \textit{High level of fault-tolerance}: Storing and processing data in a single DC may not be fault-tolerant when the DC crashes. The new system should also allow data replication from one DC to different trusted DCs, resulting in a higher level of fault-tolerance~\cite{DBLP:conf/sigmod/ZakharyNAA16}. (Note that this property is somewhat in conflict with the privacy issues. These types of systems will be reviewed under the category of \textit{frameworks for user-located geo-distributed big-data} in \S\ref{subsubsec:explicit distributed data}.)
\end{itemize}

\noindent\textbf{Advantages of geo-distributed data processing.} The main advantages of geo-distributed big-data processing are given in~\cite{DBLP:conf/sc/GhitYE12} and listed below:
\begin{itemize}[nolistsep,noitemsep,leftmargin=0.14in]
  \item A geo-distributed Hadoop/Spark-based system can perform data processing across nodes of multiple clusters while the standard Hadoop/Spark and their variants cannot process data at multiple clusters~\cite{DBLP:conf/sc/GhitYE12}.

  \item More flexible services, \textit{e}.\textit{g}., resource sharing, load balancing, fault-tolerance, performance isolation, data isolation, and version isolation, can be achieved when a cluster is a part of a geo-distributed cluster~\cite{DBLP:conf/IEEEcloud/ZhangLLZSMGA14,g-hadoop}.

  \item A cluster can be scaled dynamically during the execution of a geo-distributed computation~\cite{DBLP:conf/sc/GhitYE12}.

  \item The computation cost can be optimized by selecting different types of virtual nodes in clouds according to the user requirement and transferring a job to multiple clouds~\cite{resilin}.
\end{itemize}

\begin{figure*}
\centering
\includegraphics[scale=0.45]{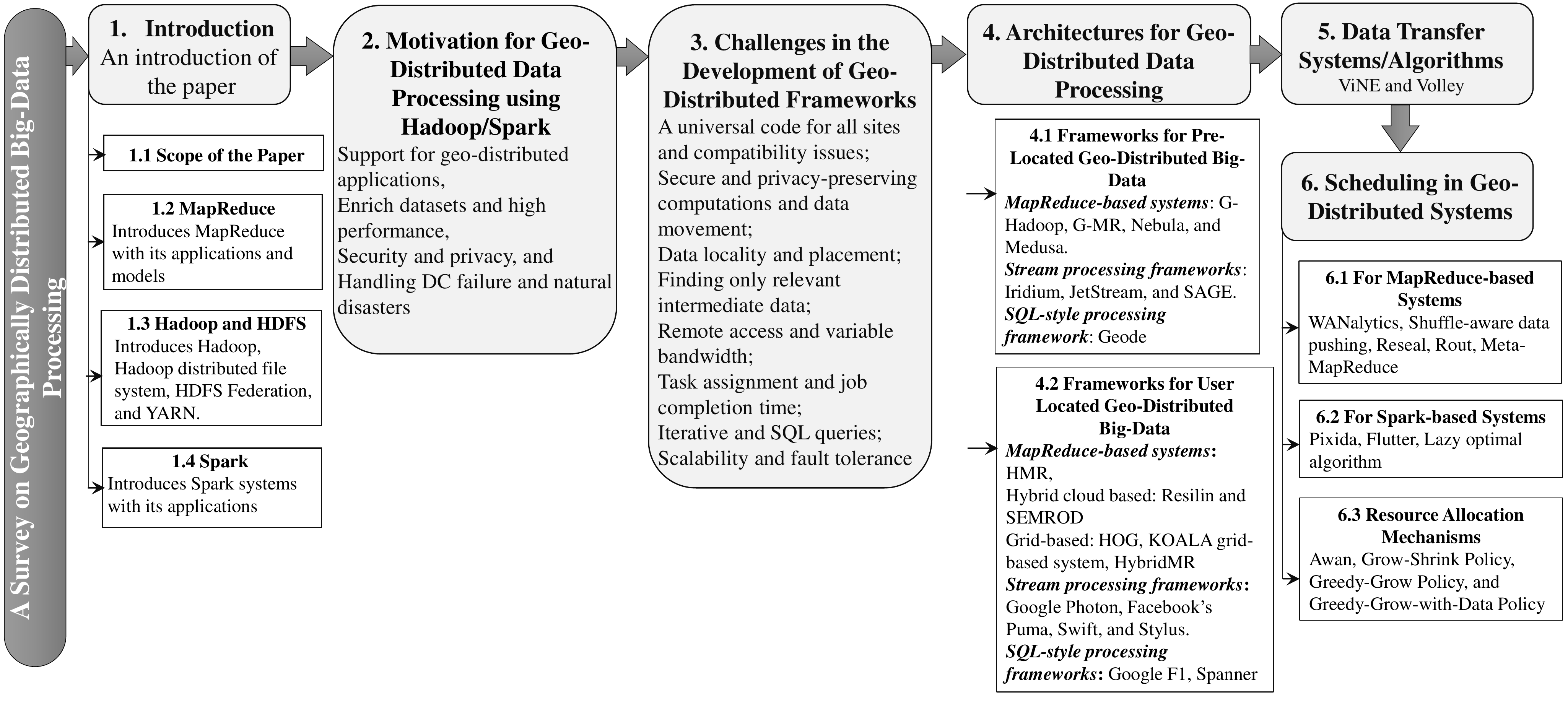}
\BBB
\BB
\caption{Schematic map of the paper.}
\label{fig:fig_box}
\BB
\end{figure*}

\BB
\subsection{Scope of the Review}
Today, big-data is a reality yielded by the distributed internet of things that constantly collect and process sensing information from remote locations. Communication and processing across different geographic areas are major resources that should be optimized. Other aspects such as regulations and privacy-preserving are also important criteria. Our paper is also motivated by these important emerging developments of big-data.

A schematic map of the paper is given in Fig.~\ref{fig:fig_box}. In this paper, we discuss design requirements, challenges, proposed frameworks, and algorithms to Hadoop-based geo-distributed data processing. It is important to emphasize that this work is not only limited to MapReduce-based batch processing geo-distributed frameworks; we will discuss architectures designed for geo-distributed streaming data (SAGE~\cite{DBLP:conf/ipps/TudoranAB13} and JetStream~\cite{DBLP:conf/nsdi/RabkinASPF14}), Spark-based systems (Iridium~\cite{akella-15}), and SQL-style processing frameworks (Geode~\cite{DBLP:conf/nsdi/VulimiriCGJPV15} and Google's Spanner~\cite{DBLP:conf/sigmod/BaconBBCDFFGJKL17}). Open issues to be considered in the future are given at the end of the paper in \S\ref{sec:Concluding Remarks}.

In this survey, we do not study techniques for multi-cloud deployment, management, and migration of virtual machines, leasing cost models, security issues in the cloud, API design, scheduling strategies for non-MapReduce jobs, and multi-cloud database systems.

\BB
\subsection{MapReduce}
\label{subsec:MapReduce}
\begin{figure}[h]
\begin{flushleft}
\centering
\includegraphics[scale=0.35]{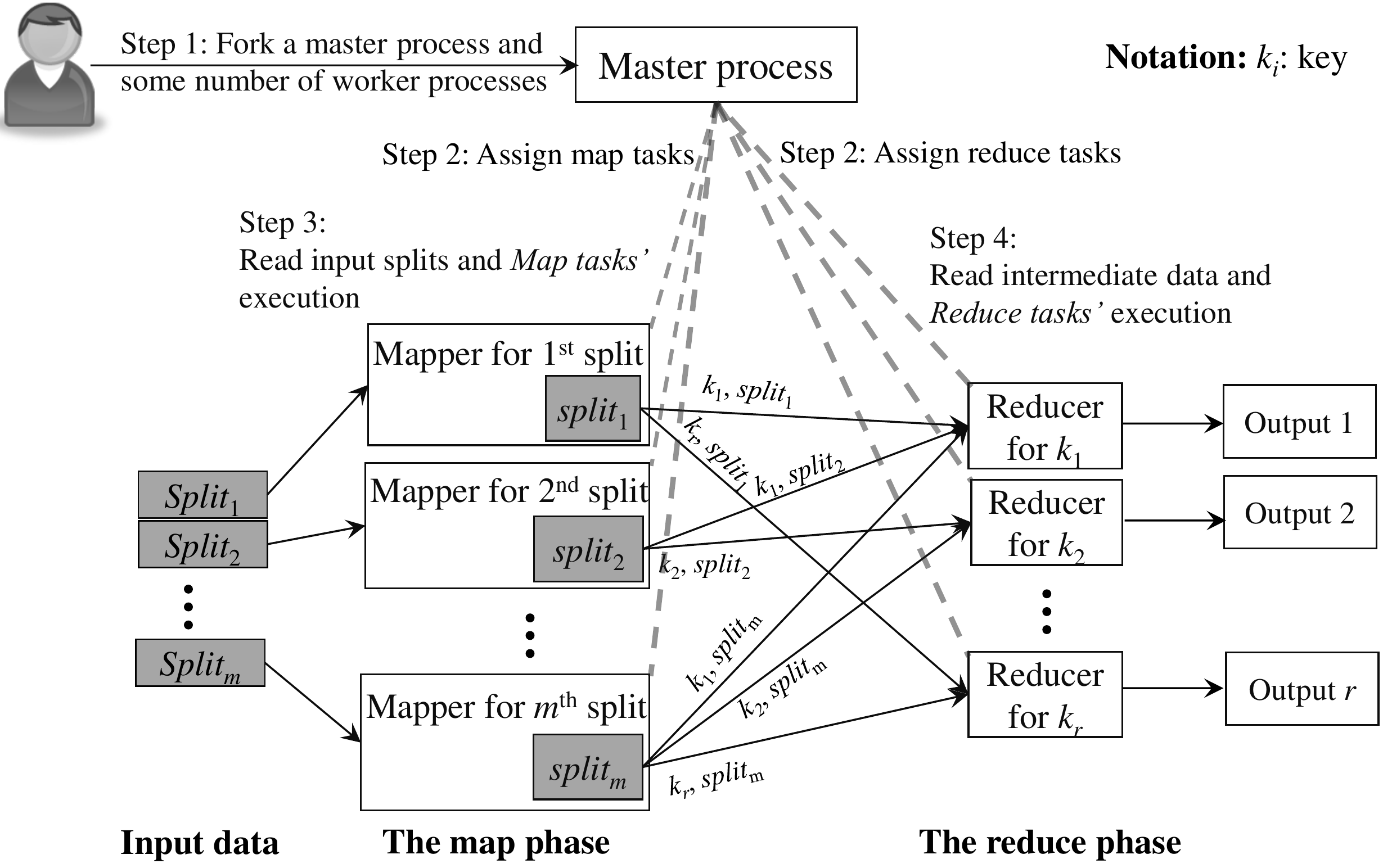}
\B
\caption{A general execution of a MapReduce algorithm.}
\label{fig:mapreduce}
\BBB\B
\end{flushleft}
\end{figure}

MapReduce~\cite{DBLP:conf/osdi/DeanG04}, introduced by Google 2004, provides parallel processing of large-scale data in a timely, failure-free, scalable, and load balance manner. MapReduce (see Fig.~\ref{fig:mapreduce}) has two phases, the {\em map phase} and the {\em reduce phase}. The given input data is processed by the map phase that applies a user-defined map function to produce intermediate data (of the form $\langle key, value \rangle$). This intermediate data is, then, processed by the reduce phase that applies a user-defined reduce function to keys and their associated values. The final output is provided by the reduce phase. A detailed description of MapReduce can be found in Chapter 2 of~\cite{DBLP:books/ullman2011}.

\smallskip \noindent\textit{Applications and models of MapReduce}. Many MapReduce applications in different areas exist. Among them: matrix multiplication~\cite{DBLP:journals/corr/abs-1204-1754}, similarity join~\cite{DBLP:conf/sigmod/VernicaCL10,DBLP:conf/icde/AfratiSMPU12}, detection of near-duplicates~\cite{DBLP:conf/www/MankuJS07}, interval join~\cite{DBLP:conf/edbt/ChawdaGNFSM14,DBLP:conf/edbt/Afrati15}, spatial join~\cite{DBLP:conf/wise/GuptaC14,DBLP:conf/sigmod/TauheedHA15}, graph processing~\cite{DBLP:conf/icde/AfratiFU13,DBLP:conf/edbt/MalhotraAS14}, pattern matching~\cite{DBLP:conf/appt/LiuJCMZ09}, data cube processing~\cite{NYBR12,DBLP:conf/sigmod/MiloA16}, skyline queries~\cite{DBLP:conf/icdt/AfratiKSU12}, $k$-nearest-neighbors finding~\cite{DBLP:conf/edbt/ZhangLJ12,DBLP:journals/pvldb/LuSCO12}, star-join~\cite{DBLP:conf/cloudi/ZhouZW13}, theta-join~\cite{DBLP:conf/sigmod/OkcanR11,DBLP:journals/pvldb/ZhangCW12}, and image-audio-video-graph processing~\cite{DBLP:journals/concurrency/YuWT0LV12}, are a few applications of MapReduce in the real world.
Some efficient MapReduce computation models for a single cloud are presented by Karloff et al.~\cite{DBLP:conf/soda/KarloffSV10}, Goodrich~\cite{DBLP:journals/corr/abs-1004-4708}, Lattanzi et al.~\cite{DBLP:conf/spaa/LattanziMSV11}, Pietracaprina et al.~\cite{DBLP:conf/ics/PietracaprinaPRSU12}, Goel and Munagala~\cite{DBLP:journals/corr/abs-1211-6526}, Ullman~\cite{DBLP:journals/crossroads/Ullman12}, Afrati et al.~\cite{DBLP:journals/pvldb/AfratiSSU13,DBLP:conf/ideas/AfratiU13,TR}, and Fish et al.~\cite{DBLP:conf/wdag/FishKLRT15}.

\BB
\subsection{Hadoop, HDFS, HDFS Federation, and YARN}
\label{subsec:Hadoop}
\noindent\textbf{Hadoop.} Apache Hadoop~\cite{apache_hadoop} is a well-known and widely used open-source software implementation of MapReduce for distributed storage and distributed processing of large-scale data on clusters of nodes. Hadoop includes three major components, as follows: (\textit{i}) Hadoop Distributed File System (HDFS)~\cite{HDFS2010}: a scalable and fault-tolerant distributed storage system, (\textit{ii}) Hadoop MapReduce, and (\textit{iii}) Hadoop Common, the common utilities, which support the other Hadoop modules.

\begin{figure}[!t]
\centering
\includegraphics[scale=0.4]{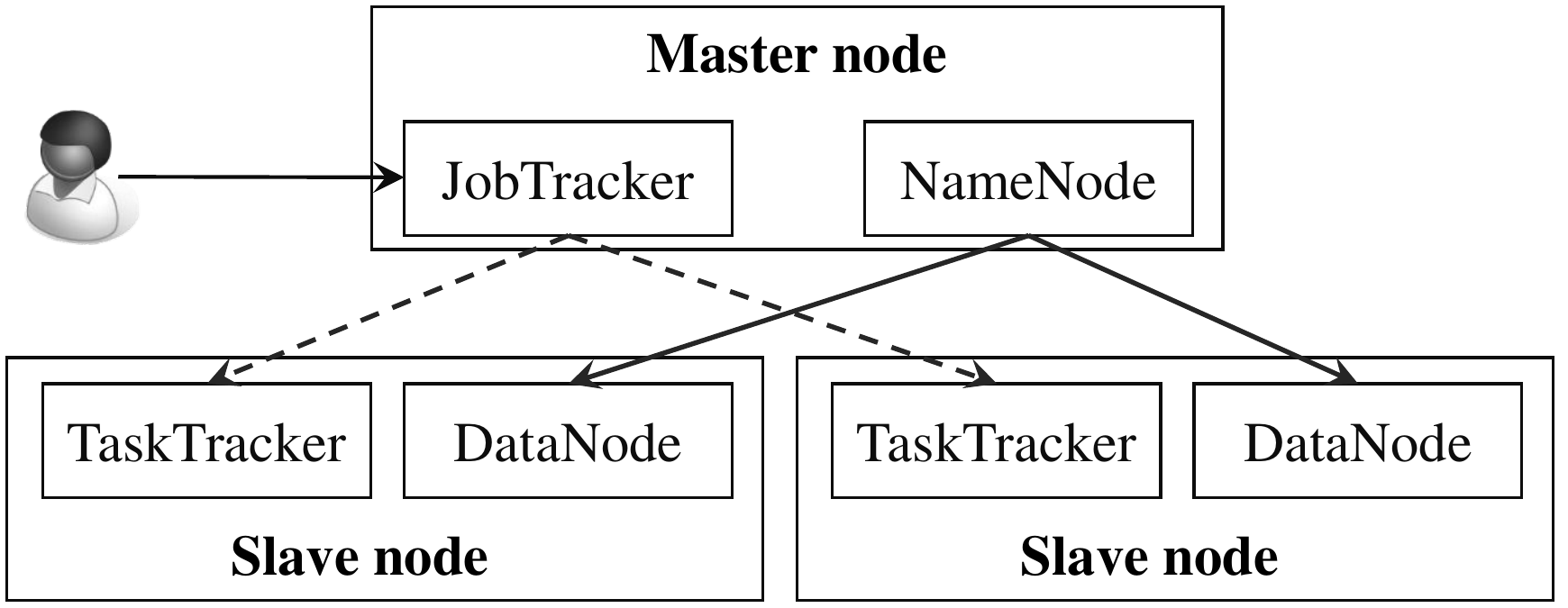}
\B
\caption{Structure of a Hadoop cluster with one master node and two slave nodes.}
\label{fig_hadoop}
\BBB
\end{figure}

Hadoop cluster consists of two types of nodes, as; (\textit{i}) a master node that executes a JobTracker and a NameNode and (\textit{ii}) several slave nodes, each slave node executes a TaskTracker and a DataNode; see Fig.~\ref{fig_hadoop}. The computing environment for a MapReduce job is provided by the JobTracker (that accepts a MapReduce job from a user and executes the job on free TaskTrackers) and TaskTrackers (produces the final output). An environment for distributed file system, called HDFS is supported by the NameNode (manages the cluster metadata and DataNodes) and DataNodes (stores data). HDFS supports \texttt{read}, \texttt{write}, and \texttt{delete} operations on files, and \texttt{create} and \texttt{delete} operations on directories. In HDFS, data is divided into small splits, called \emph{blocks}, (64MB and 128MB are most commonly used sizes). Each block is independently replicated to multiple DataNodes, and block replicas are processed by mappers and reducers. More details about Hadoop and HDFS may be found in Chapter 2 of~\cite{lin2010data}.

\smallskip\noindent\textbf{HDFS Federation.} In the standard HDFS, there is only one NameNode, which is a single point of failure. HDFS Federation~\cite{hdfs-f} overcomes this limitation of HDFS by adding multiple NameNodes that are independent and do not require coordination with each other. DataNodes store data, and in addition, each DataNode registers with all the NameNodes. In this manner, HDFS Federation creates a large virtual cluster that increases performance, turns NameNode to be fault-tolerant, and provides multiple isolated jobs' execution framework.

\smallskip\noindent\textbf{YARN Architecture.} YARN~\cite{yarn} is the latest version of Hadoop-2.7.1 and partitions the two major functionalities of the JobTracker of the previous Hadoop, \textit{i}.\textit{e}., resource management and job scheduling monitoring, into separate daemons, called a global ResourceManager daemon and a per-application ApplicationMaster daemon. Details about YARN may be found in~\cite{murthy2013apache}.

The ResourceManager is responsible for dividing the cluster's resources among all the applications running in the system. The ApplicationMaster is an application-specific entity, negotiates resources from the ResourceManager. The NodeManager is a per-node daemon, which is responsible for launching the application's containers, monitoring their resource usage (CPU, memory etc.), and reporting back to the ResourceManager. A container represents a collection of physical resources.


\BB
\subsection{Spark}
\label{subsec:spark}
Apache Spark is a cluster computing platform that extends MapReduce-style processing for efficiently supporting more types of fast and real-time computations, interactive queries, and stream processing. The major difference between Spark and Hadoop lies in the processing style, where MapReduce stores outputs of each iteration in the disk while Spark stores data in the main memory, and hence, supports fast processing. Spark also supports Hadoop, and hence, it can access any Hadoop data sources. Spark Core contains task scheduling, memory management, fault recovery, interacting with storage systems, and defines \textit{resilient distributed datasets} (RDDs). RDDs are main programming abstraction and represent a collection of distributed items across many computing nodes that can execute a computation. Spark supports several programming languages such as Python, Java, and Scala. Details about Spark may be found in~\cite{spark_book}.

Matrix computations~\cite{DBLP:conf/kdd/ZadehMUYPVSSZ16}, machine learning~\cite{DBLP:journals/corr/MengBYSVLFTAOXX15}, graph processing~\cite{DBLP:conf/sigmod/XinGFS13}, iterative queries~\cite{spark_book}, and stream processing~\cite{DBLP:conf/hotcloud/ZahariaDLSS12} are a few popular examples of computational fields where Spark is commonly used. Apache Flink~\cite{flink}, Apache Ignite~\cite{Ignite}, Apache Storm~\cite{Storm}, and Twitter Heron~\cite{DBLP:conf/sigmod/KulkarniBFKKMPR15} are other stream processing frameworks.

\BB
\section{Motivations and Needs for Geo-Distributed Big-Data Processing using Hadoop or Spark}
\label{section:Requirements of Distributed Hadoop}

We list four major motivational points behind the design of a geo-distributed big-data processing framework, as follows:

\smallskip\noindent\textbf{Support for geo-distributed applications.} As we mentioned in \S\ref{sec:introduction}, a lot of applications generate data at geo-distributed locations or DCs. On the one hand, genomic and biological data, activity, session and server logs, and performance counters are expanding geographically much faster than inter-DC bandwidth; hence, such entire datasets cannot be efficiently transferred to a single location~\cite{DBLP:conf/cidr/VulimiriCGKV15,akella-15,DBLP:conf/icdcs/JayalathE13,canada}. On the other hand, analysis and manipulation operations do not require an entire dataset from each location in providing the final output. Thus, there is a need of on-site big-data processing frameworks that can send only the desired inputs (after processing data at each site) to a single location for providing the final outputs under legal constraints of an organization.

\smallskip\noindent\textbf{Enrich datasets and high performance.} Currently, a data-intensive application produces, manipulates, or analyzes data of size Petabytes or more. Sharing such a huge amount of data across the globe enriches datasets and helps several communities in finding recent trends, new types of laws, regulations, and networking constraints~\cite{canada}. In contrast, the data processing using MapReduce is dominated by the communication between the map phase and the reduce phase, where several replicas of an identical data are transferred to the reduce phase for obtaining final outputs~\cite{DBLP:journals/pvldb/AfratiSSU13,TR,metamr_sss-15}. However, sometimes none of the replicated partial outputs provides the final outputs.

For example, we want to execute a top-k query on $n$ locations that have their own data. In this case, a trivial solution is to send the whole data from all the $n$ locations to a single location. In contrast, we can execute a top-k query on each $n$ location and send only top-k results from each $n$ locations to a single location that can find the final top-k answers. One more example, in case of equijoin of two relations, say $X(A,B)$ and $Y(B,C)$, where $X$ and $Y$ are located in different sites, if there is no tuple containing a joining value, say $b_1$, in the relation $Y$, then it is worthless to transfer all the tuples of the relation $X$ having $b_1$ to the location of the relation $Y$.

Therefore, it is clear and trivial that one can solve geo-applications by moving an entire dataset to a single location; however, it will increase the network load, job completion time, space requirement at the single site, and decrease the performance of the system. Thus, the development of efficient geo-distributed Hadoop or Spark-based frameworks transparent to users is needed~\cite{akella-15}.

\smallskip\noindent\textbf{Providing geo-security and geo-privacy mechanisms.}
The development of a geo-distributed framework, inherently, requires a secure and privacy-preserving mechanism for data transfer and computation execution. The design of geo-security and geo-privacy mechanisms will help in geo-frameworks, also in a single cloud computing platform, where data and computation locations are identical.

Here, we explain why do the current security and privacy mechanisms fail in a geo-distributed framework. Data may be classified as public, sensitive, or confidential with special handling~\cite{zerbegeospatial}. Executing a geo-application on public data of an organization may breach the security and privacy of sensitive or confidential data, since a geo-application may attempt to scan the entire dataset. For example, executing a geo-application on a health data of a country allows data movement within the country; however, the same data may not be allowed to be accessed by a geo-application that is moving the data outside the country. Another example of privacy breaking that can occur when the public data of the geo-locations of a person is associated with the disease data which is sensitive~\cite{zerbegeospatial}. In a similar manner, the data confidentiality is vulnerable in a geo-computation.

Thus, the security and privacy in a geo-computation depend on several factors, \textit{e}.\textit{g}., organizations that are asking for data, the organizations' location, and the scope of the computation. In addition, if on-site data processing is allowed, an organization wishes to ensure the security and privacy of their output data according to their policies, during the transmission and computations at other sites, resulting in no malicious activities on the data~\cite{DBLP:conf/cidr/VulimiriCGKV15}. Therefore, the design of geo-secure and geo-privacy mechanisms is required, thus, maintaining data security and privacy during a geo-distributed computation. However, creating a secure and privacy-preserving geo-distributed framework raises several challenges, which we discuss in the next section.

\smallskip\noindent\textbf{Handling DC failure and natural disasters.} The classical Hadoop was designed to prevent a job failure due to disk, node, or rack failure by replicating the data along with the job to multiple nodes and racks within a DC. A single DC failure/outage is not very common; however, if it does happen, then it leads to severe obstacles. For example, in the month of May 2017, due to the power outage, British Airways DC was crashed, and that leads to catastrophic impacts.\footnote{\url{https://www.theguardian.com/business/2017/may/30/british-airways-it-failure-experts-doubt-power-surge-claim}} In order to handle DC failure, replication of the data with the jobs to different DCs (possibly outside the country) is expected to be a trivial solution. However, such a geo-replication requires us to redesign new Hadoop/Spark-based systems that can work over different DCs transparent to any failure. We will study some frameworks supporting geo-replication in \S\ref{subsubsec:explicit distributed data}.


\BB
\section{Challenges in the Development of Geo-Distributed Hadoop or Spark based Frameworks}
\label{sec:Challenges in the Development of Geo-Distributed Framework}
The existence of big-data, on one hand, requires the design of a fault-tolerant and computation efficient framework, and Hadoop, Spark, or similar frameworks satisfy these requirements. On the other hand, globally distributed big-databases --- as opposed to traditional parallel databases, cluster computations, and file-sharing within a DC --- introduce new research challenges in different domains, as follows: (\textit{i}) the database domain has new challenges such as query planning, data locality, replication, query execution, cost estimation, and the final output generation; (\textit{ii}) the wide area network domain has new challenges such as bandwidth constraints and data movement~\cite{DBLP:conf/nsdi/VulimiriCGJPV15}. In addition, geo-distributed data processing using the current frameworks inherits some old challenges such as location transparency, (\textit{i}.\textit{e}., a user will receive a correct output regardless of the data location), and local autonomy, (\textit{i}.\textit{e}., the capability to administer a local database and to operate independently when connections to other nodes have failed)~\cite{DBLP:conf/IEEEcloud/ZhangLLZSMGA14,DBLP:conf/nsdi/VulimiriCGJPV15}.

In this section, we describe new challenges in the context of geo-distributed big-data processing using Hadoop or Spark-based systems. After each challenge, we give references to solutions to the challenge, which are described later in this paper.

\smallskip\noindent\textbf{A universal code for all sites and compatibility issues.} In the present time, several big-data processing frameworks, languages, and mechanisms are proposed; for example, Hadoop, Yarn, Spark, Hive~\cite{DBLP:journals/pvldb/ThusooSJSCALWM09,DBLP:conf/sigmod/HuaiCGHHOPYL014}, Pig Latin~\cite{DBLP:conf/sigmod/OlstonRSKT08}, Dryad, Spark SQL~\cite{DBLP:conf/sigmod/ArmbrustXLHLBMK15}, etc. In addition, different big-data and metadata storages, like HDFS, Gfarm file system~\cite{tatebe2010gfarm}, GridDB~\cite{DBLP:conf/sigmod/LiuFP03}, MongoDB~\cite{banker2011mongodb}, HBase~\cite{george2011hbase}, etc., are available. These databases have non-identical data formats, APIs, storage policies, privacy concerns for storing and retrieving data, network dynamics, and access control~\cite{DBLP:journals/pvldb/CooperRSSBJPWY08,DBLP:journals/pvldb/GuptaYGKCLWDKABHCSJSGVA14,DBLP:conf/usenix/BronsonACCDDFGKLMPPSV13}.

Moreover, different sites may have different types of regulations for exporting data, operating systems, availability of data, services, security checks, resources, cost, and software implementations. Sometimes, simultaneous usages of multiple frameworks improve utilization and allow applications to share access to large datasets~\cite{DBLP:conf/nsdi/HindmanKZGJKSS10}. However, the existence of different frameworks at different locations poses additional challenges such as different scheduling needs, programming models, communication patterns, task dependencies, data placement, and different APIs.

Hence, according to a client perspective, it is not desirable to write code for different frameworks and different data formats. For example, if there are two sites with HDFS and Gfarm file system, then the data retrieval code is not identical and the user has to write two different codes for retrieving data. In this scenario, it is required that a universal code will work at all the locations without modifying their data format and processing frameworks. It may require an interpreter that converts a user code according to the requirement of different frameworks. However, the use of an interpreter puts some additional challenges such as how does a system follow the inherent properties, \textit{e}.\textit{g}., massive parallelism and fault-tolerance. It should be noted that user-defined compatibility tasks may slow down the overall system performance~\cite{DBLP:conf/icdcs/JayalathE13,DBLP:journals/pc/HawickCJ03}. Unfortunately, there is not a single geo-distributed system that can solve this challenge, to the best of our knowledge.

\smallskip\textit{Solutions}: Mesos~\cite{DBLP:conf/nsdi/HindmanKZGJKSS10} provides a solution to the above-mentioned requirements to some extents. Twitter's Summingbird~\cite{DBLP:journals/pvldb/BoykinROL14} integrates batch and stream processing within a single DC. Recently, BigDAWG~\cite{DBLP:conf/hpec/GadepallyCDEHKM16} and Rheem~\cite{DBLP:conf/edbt/AgrawalCEKOPQ0Z16} are two new systems that are focusing on compatibility issues in a single DC. In short, BigDAWG~\cite{DBLP:conf/hpec/GadepallyCDEHKM16} and Rheem~\cite{DBLP:conf/edbt/AgrawalCEKOPQ0Z16} are trying to achieve platform-independent processing, multi-platform task execution, exploit complete processing capabilities of underlying systems, and data processing abstraction (in a single DC).
Since Mesos~\cite{DBLP:conf/nsdi/HindmanKZGJKSS10}, Summingbird~\cite{DBLP:journals/pvldb/BoykinROL14}, BigDAWG~\cite{DBLP:conf/hpec/GadepallyCDEHKM16}, and Rheem~\cite{DBLP:conf/edbt/AgrawalCEKOPQ0Z16} deals with processing in a single DC, we do not study these systems in this survey.

Awan~\cite{DBLP:conf/ic2e/JonathanCW16} (\S\ref{subsec:Resource Allocation Mechanism for Distributed Hadoop}) is a system that allocates resources to different geo-distributed frameworks. However, Awan does not provide universal compatibilities to the existing systems. Also, during our investigation, we did not find any system that can handle above-mentioned compatibility issues in the geo-distributed settings.

\smallskip\noindent\textbf{Secure and privacy-preserving computations and data movement.} Geo-distributed applications are increasing day-by-day, resulting in an increasing number of challenges in maintaining fine and coarse grain security and privacy of data or computations. The classical MapReduce does not support the security and privacy of data or computations within a single public cloud. But even if the security and privacy in a single cloud is preserved, it is still a major challenge in ensuring the security and privacy of data or computations in geo-distributed big-data processing. A survey on security and privacy in the standard MapReduce may be found in~\cite{ehud-security-privacy-mr}.

The security analysis also requires risk management. Applying complex security mechanisms on a geo-computation without considering risk management may harm the computation and system performance, while it is not required to implement rigorous security systems. Hence, there is a need to consider risk when designing security and privacy mechanisms for geo-computations~\cite{white-house,DBLP:conf/infocom/HuLL16}. The privacy of an identical type of data (\textit{e}.\textit{g}., health-care data) may not be treated the same when implemented in different countries. Hence, there are several issues to be addressed while designing a secure geo-distributed framework, as follows: how to trust the data received from a site, how to ensure that the data is be transferred in a secure and private manner, how to build trust among sites, how to execute computations in each site in a privacy-preserving manner, how to pre-inspect programs utilizing the data, how to ensure usage of the data, and how to allow a computation execution while maintaining fine-grained security features such as authentication, authorization, and access control~\cite{DBLP:journals/internet/KeaheyTMF09}.

\smallskip\textit{Solutions}: G-Hadoop~\cite{g-hadoop} (\S\ref{subsubsec:Implicit distributed data}) provides an authentication mechanism, and ViNe~\cite{DBLP:journals/internet/KeaheyTMF09} (\S\ref{subsec:Data Transfer Systems}) provides an end-to-end data security. However, currently, we are not aware of any complete solution for security and privacy in the context of geo-distributed Hadoop/Spark jobs.

\smallskip\noindent\textbf{Data locality and placement.} In the context of geo-data processing, data locality refers to data processing at the same site or nearby sites where the data is located~\cite{DBLP:conf/hotos/RabkinASPF13,DBLP:conf/infocom/HuLL16}. However, the current Hadoop/Spark/SQL-style based geo-distributed data processing systems are designed on the principle of data pulling from all the locations to a single location, and hence, they do not regard data locality~\cite{DBLP:conf/nsdi/VulimiriCGJPV15}. In addition, due to a huge amount of raw data generated at different sites, it is challenging to send the whole dataset to a single location; hence, the design and development of systems that take the data locality into account are crucial for optimizing the system performance.

In contrast, sometimes a framework regarding the data locality does not work well in terms of performance and cost~\cite{DBLP:conf/hpdc/OhRCW15,DBLP:conf/sigmod/ZakharyNAA16,DBLP:journals/tc/ChenPL16} due to the limited number of resources or slow inter-DC connections. Hence, it may be required to access/process data in nearby DCs, which may be faster than the local access. Thus, we find a challenge in designing a system for accessing local or remote (nearby) data, leading to optimized job performance.

\smallskip\textit{Solutions}: G-Hadoop~\cite{g-hadoop} (\S\ref{subsubsec:Implicit distributed data}), GMR~\cite{DBLP:journals/tc/JayalathSE14} (\S\ref{subsubsec:Implicit distributed data}), Nebula~\cite{DBLP:conf/ic2e/RydenOCW14} (\S\ref{subsubsec:Implicit distributed data}), Iridium~\cite{akella-15} (\S\ref{subsubsec:A frameworks for implicit geo-distributed streaming data}), and JetStream~\cite{DBLP:conf/nsdi/RabkinASPF14} (\S\ref{subsubsec:A frameworks for implicit geo-distributed streaming data}).

\smallskip\noindent\textbf{Finding only relevant intermediate data.} A system built on the ``data locality'' principle processes data at their sites and provides intermediate data. However, sometimes, the complete intermediate data at a site do not participate in the final output, and hence, it necessitates to find only relevant intermediate data. We emphasize that the concepts of the data locality and relevant intermediate data finding are different.

A computation can be characterized by two parameters: (\textit{i}) the amount of input data (in bits) at a data source, and (\textit{ii}) the expansion factor, $e$, which shows a ratio of the output size to the input size~\cite{cardosa2011exploring,heintz2015end,DBLP:conf/IEEEcloud/ZhangLLZSMGA14,akella-15}. Based on the expansion factor, a computation can be of three types, as follows: (\textit{i}) $e\gg 1$: the output size is much larger than the input size, \textit{e}.\textit{g}., join of relations; (\textit{ii}) $e=1$: the output size is of the same size as the input size, \textit{e}.\textit{g}. outputs of a sorting algorithm; (\textit{iii}) $e\ll 1$: the output size is less than the input size, \textit{e}.\textit{g}., word count. When dealing with a geo-distributed computation, it is not efficient to move all the data when $e\gg 1$ and only some parts of that data participate in the final outputs~\cite{DBLP:conf/icdcs/JayalathE13,metamr_sss-15}. For example, in the first, second, and third types of computations, if the computation performs a joining of relations while most of the tuples of a relation do not join with any other relations at the other sites, a global sorting only on selected intermediate sorted outputs, and a frequency-count of some words, respectively, then we need to find only relevant intermediate data.

These challenges motivate us to find only relevant intermediate data at different locations before obtaining the final output. In addition, it is also required to prioritize data considering dynamic requirements, resources, and usefulness of data before moving data~\cite{DBLP:conf/hotos/RabkinASPF13}.

\smallskip\textit{Solutions}: Iridium~\cite{akella-15} (\S\ref{subsubsec:A frameworks for implicit geo-distributed streaming data}), Geode~\cite{DBLP:conf/nsdi/VulimiriCGJPV15} (using difference finding, \S\ref{subsubsec:A framework for implicit geo-distributed data processing using SQL}), and Meta-MapReduce~\cite{metamr_sss-15} (\S\ref{subsec:Scheduling for Geo-distributed MapReduce-based Systems}).

\smallskip\noindent\textbf{Remote access and variable bandwidth.} The cost of a geo-distributed data processing is dependent on remote accesses and the network bandwidth, and as the amount of inter-DC data movement increases, the job cost also increases~\cite{DBLP:conf/infocom/HuLL16,g-cut}. In addition, we may connect DCs with a low-latency and high-bandwidth interconnects for fine-grain data exchanges and that results in a very high cost. Hence, an intelligent remote data access mechanism is required for fetching only the desired data. Moreover, frequently accessed data during the execution of similar types of jobs can be placed in some specific DCs to reduce the communication~\cite{DBLP:conf/cidr/VulimiriCGKV15}.

The limited bandwidth constraint, thus, motivates to design efficient distributed and reliable systems/algorithms for collecting and moving data among DCs while minimizing remote access, resulting in lower job cost and latency~\cite{heintz2015end,DBLP:conf/ipps/TandonCW13}. In addition, the system must adjust the network bandwidth dynamically; hence, the system can drop some data without affecting data quality significantly~\cite{DBLP:conf/nsdi/RabkinASPF14}. In other words, there is a need of an algorithm that will know the global view of the system (consisting of the bandwidth, data at each DC, and distance to other DCs).

Serving data transfer as best-effort or real—time is also a challenge in geo-computations. On the one hand, if one serves best-effort data transfer, then the final output will be delayed, and it requires a significant amount of the network bandwidth at a time. On the other hand, if we transfer data in real-time, then the user will get real-time results at the cost of endless use of the network bandwidth~\cite{DBLP:conf/sosp/OusterhoutWZS13}.

\smallskip\textit{Solutions}: Iridium~\cite{akella-15} (\S\ref{subsubsec:A frameworks for implicit geo-distributed streaming data}), Pixida~\cite{DBLP:journals/pvldb/KloudasRPM15} (\S\ref{subsec:Scheduling for Geo-distributed Spark-based Systems}), Lazy optimal algorithm~\cite{DBLP:conf/hpdc/HeintzCS15} (\S\ref{subsec:Scheduling for Geo-distributed Spark-based Systems}), JetStream~\cite{DBLP:conf/nsdi/RabkinASPF14} (\S\ref{subsubsec:A frameworks for implicit geo-distributed streaming data}), Volley~\cite{DBLP:conf/nsdi/AgarwalDJSW10} (\S\ref{subsec:Scheduling for Geo-distributed MapReduce-based Systems}), Rout~\cite{DBLP:conf/icdcs/JayalathE13} (\S\ref{subsec:Scheduling for Geo-distributed MapReduce-based Systems}), and Meta-MapReduce~\cite{metamr_sss-15} (\S\ref{subsec:Scheduling for Geo-distributed MapReduce-based Systems}).

\smallskip\noindent\textbf{Task assignment and job completion time.} The job completion time of a geo-distributed computation is dependent on several factors, as follows: (\textit{i}) data locality, (\textit{ii}) the amount of intermediate data, (\textit{iii}) selection of a DC for the final task --- it is required to assign the final task to a DC that has a major portion of data that participate in the final outputs, resulting in fast job completion and reduced data transfer time~\cite{DBLP:journals/sigmetrics/GadreRP11,DBLP:conf/ccgrid/TudoranCWBA14}, and (\textit{iv}) the inter-DC and intra-DC bandwidth. The authors~\cite{DBLP:conf/IEEEcloud/YazdanovGF15} showed that an execution of a MapReduce job on a network-aware vs. network-unaware scheme significantly impacts the job completion time.

The challenge comes in finding a straggler process. In a locally distributed computation, we can find straggler processes and perform a speculative execution for fast job completion time. However, these strategies do not help in a geo-distributed data processing, because of different amount of data in DCs and different bandwidth~\cite{DBLP:conf/hotos/RabkinASPF13}.

In addition, the problem of straggler processes cannot be removed by applying offline task optimization placement algorithms, since they rely on a priori knowledge of task execution time and inter-DC transfer time, which both are unknown in geo-distributed data processing~\cite{DBLP:conf/infocom/HuLL16}.

\smallskip\textit{Solutions}: Iridium~\cite{akella-15}, Joint optimization of task assignment, data placement, and routing in geo-distributed DCs~\cite{DBLP:journals/tetc/GuZLG14}, Reseal~\cite{DBLP:conf/ipps/KettimuthuASF16} (\S\ref{subsec:Scheduling for Geo-distributed MapReduce-based Systems}), Tudoran et al.~\cite{DBLP:conf/ccgrid/TudoranCWBA14} (\S\ref{subsec:Data Transfer Systems}) and Gadre et al.~\cite{DBLP:conf/cac/GadreRMP13} (\S\ref{subsec:Data Transfer Systems}), and Flutter~\cite{DBLP:conf/infocom/HuLL16} (\S\ref{subsec:Scheduling for Geo-distributed Spark-based Systems}).



\smallskip\noindent\textbf{Iterative and SQL queries.} The standard MapReduce was not developed for supporting iterative and a wide range of SQL queries. Later, Hive, Pig Latin, and Spark SQL were developed for supporting SQL-style queries. However, all these languages are designed for processing in-home/local data. Since relational algebra is a basis of several different operations, it is required to develop a geo-distributed query language regarding data locality and the network bandwidth. The new type of query language must also deal with some additional challenges such as geo-distributed query optimization, geo-distributed query execution plan, geo-distributed indexing, and geo-distributed caching. The problem of joining of multiple tables that are located at different locations is also not trivial. In this case, moving an entire table from one location to the location of the other table is naive yet cumbersome, because of network bandwidth, time, and cost. Hence, we see the joining operation in geo-distributed settings is a major challenge. The joining operation gets more complicated in the case of streaming of tables where a window-based join does not work~\cite{DBLP:conf/sigmod/AnanthanarayananBDGJQRRSV13} because the joining values of multiple tables may not \emph{synchronously} arrive at an identical time window, thereby leading to missing outputs.

Processing iterative queries on the classical Hadoop was a cumbersome task due to disk-based storage after each iteration (as we mentioned the difference between Hadoop and Spark in \S\ref{subsec:spark}). However, Spark can efficiently process iterative queries due to \emph{in-memory} processing. Processing iterative queries in a geo-computation requires us to find solutions to store intermediate results in the context of an iterative query.

\smallskip\textit{Solutions}: Geode~\cite{DBLP:conf/nsdi/VulimiriCGJPV15} (\S\ref{subsubsec:A framework for implicit geo-distributed data processing using SQL}) provides a solution to execute geo-distributed SQL queries. Google's F1~\cite{DBLP:journals/pvldb/ShuteVSHWROLMECRSA13} and Spanner~\cite{DBLP:conf/sigmod/BaconBBCDFFGJKL17} (\S\ref{subsubsec:SQL-style processing framework for user-located geo-distributed data}) are two SQL processing systems. There are some other systems~\cite{DBLP:journals/corr/abs-1303-3517,DBLP:journals/access/ZhangC14} for machine learning based on iterative Hadoop/Spark processing. However, in this paper, we are not covering any paper regarding machine learning using Hadoop/Spark.

\smallskip\noindent\textbf{Scalability and fault-tolerance.} Hadoop, Yarn, Spark, and similar big-data processing frameworks are scalable and fault-tolerant as compared to parallel computing, cluster computing, and distributed databases. Because of these two features, several organizations and researchers use these systems daily for big-data processing. Hence, it is an inherent challenge to design a new fault-tolerant geo-distributed framework so that the failure of the whole/partial DC does not lead to the failure of other DCs and also scalable in terms of adding or removing different DCs, computing nodes, resources, and software implementations~\cite{DBLP:conf/icdcs/JayalathE13,DBLP:conf/sigmetrics/PotharajuJ13,hog}.

\smallskip\textit{Solutions}: Medusa~\cite{DBLP:conf/ccgrid/Costa0RC16} (\S\ref{subsubsec:Implicit distributed data}), Resilin~\cite{resilin} (\S\ref{subsubsec:explicit distributed data}), HOG~\cite{hog} (\S\ref{subsubsec:explicit distributed data}), and KOALA-grid-based system~\cite{DBLP:conf/sc/GhitYE12} (\S\ref{subsubsec:explicit distributed data}).




The above-mentioned issues will naturally impact on the design and division of functionality of different components of a framework, which are located at non-identical locations.

\BB
\section{Architectures for Geo-Distributed Big-Data Processing}
\label{sec:Architectures for Distributed Hadoop}
In this section, we review several geo-distributed big-data processing frameworks and algorithms under two categories, as follows:
\begin{table*}[h!]
\begin{center}
\bgroup
\def\arraystretch{1.105}
\begin{tabular}{|p{5.5cm}|p{1cm}|l|l|l|l|l|l|l|l|l|l|l|}\hline


\begin{sideways}Frameworks/Protocols\end{sideways} & \begin{sideways}Data distribution\end{sideways} & \begin{sideways}Data processing \end{sideways} & \begin{sideways}Security and privacy \end{sideways}& \begin{sideways}Secure data movement\end{sideways} & \begin{sideways}Optimized paths among DCs\end{sideways} & \begin{sideways}Resource management\end{sideways} & \begin{sideways}Data locality\end{sideways} & \begin{sideways}Relevant intermediate data finding\end{sideways} & \begin{sideways}Bandwidth consideration\end{sideways} & \begin{sideways}Scalability \end{sideways}& \begin{sideways}SQL-support \end{sideways} & \begin{sideways}Result caching\end{sideways}\\\hline

\multicolumn{13}{|c|}{Geo-distributed batch processing MapReduce-based systems for pre-located geo-distributed data (Section~\ref{subsubsec:Implicit distributed data})}\\\hline

G-Hadoop~\cite{g-hadoop} & P \& U & \checkmark &\checkmark$^p$ & & & &\checkmark&\checkmark&&&& \\\hline

G-MR~\cite{DBLP:journals/tc/JayalathSE14} & P & \checkmark& & & \checkmark &\checkmark &\checkmark&\checkmark&&&&\\\hline

Nebula~\cite{DBLP:conf/ic2e/RydenOCW14} & P & \checkmark& & &  &    &\checkmark&\checkmark&&&&\\\hline

Medusa~\cite{DBLP:conf/ccgrid/Costa0RC16} & P \& U & \checkmark& & &  &    &&&\checkmark&&&\\\hline

\multicolumn{13}{|c|}{Geo-distributed stream processing frameworks for pre-located geo-distributed data (Section~\ref{subsubsec:A frameworks for implicit geo-distributed streaming data})}\\\hline

Iridium~\cite{akella-15} & P &\checkmark & & & \checkmark &     &\checkmark&\checkmark&\checkmark&&&\\\hline

JetStream~\cite{DBLP:conf/nsdi/RabkinASPF14} & P & & & &  &    &\checkmark&&\checkmark&&&\\\hline

SAGE~\cite{DBLP:conf/ipps/TudoranAB13} & P & \checkmark & & & & \checkmark   &&&&&&\\\hline

\multicolumn{13}{|c|}{SQL-style processing framework for pre-located geo-distributed data (Section~\ref{subsubsec:A framework for implicit geo-distributed data processing using SQL})}\\\hline

Geode~\cite{DBLP:conf/nsdi/VulimiriCGJPV15} & P &\checkmark & & &  &   &\checkmark&\checkmark&\checkmark&&\checkmark &\checkmark  \\\hline

\multicolumn{13}{|c|}{Geo-distributed batch processing MapReduce-based systems for user-located geo-distributed data (Section~\ref{subsubsec:Geo-distributed batch processing MapReduce-based systems for user located geo-distributed data})}\\\hline

HMR~\cite{hmr} &U & \checkmark & & & &     &&&&&&\\\hline

Resilin~\cite{resilin} & U & \checkmark & & & & \checkmark     &&&&\checkmark&&\\\hline

SEMROD~\cite{DBLP:conf/sigmod/OktayMKK15} & U & \checkmark& \checkmark & & & &\checkmark&&&\checkmark & & \\\hline

HOG~\cite{hog} & U & \checkmark & & & & \checkmark &&&&\checkmark&&   \\\hline

KOALA grid-based system~\cite{DBLP:conf/sc/GhitYE12} & U & \checkmark & & & & \checkmark    &&&&\checkmark&& \\\hline

HybridMR~\cite{CPE15} & U & \checkmark& & & & &&&&\checkmark & & \\\hline

\multicolumn{13}{|c|}{Geo-distributed stream processing frameworks for user-located geo-distributed data (Section~\ref{subsubsec:Geo-distributed stream processing frameworks for user-located geo-distributed data})}\\\hline

Photon~\cite{DBLP:conf/sigmod/AnanthanarayananBDGJQRRSV13} & P \& U & \checkmark& & & &\checkmark &\checkmark&\checkmark&&\checkmark & & \\\hline

\multicolumn{13}{|c|}{SQL-style processing framework for user-located geo-distributed data (Section~\ref{subsubsec:SQL-style processing framework for user-located geo-distributed data})}\\\hline

Spanner~\cite{DBLP:conf/sigmod/BaconBBCDFFGJKL17} & P \& U & \checkmark& & & &\checkmark &\checkmark&\checkmark&&\checkmark &\checkmark & \\\hline

\multicolumn{13}{|c|}{Data transfer systems/algorithms (Section~\ref{subsec:Data Transfer Systems})}\\\hline
Tudoran et
al.~\cite{DBLP:conf/ccgrid/TudoranCWBA14}, Gadre et al.~\cite{DBLP:conf/cac/GadreRMP13}  & P \& U & & & & & \checkmark     &&&&&&\\\hline

Volley~\cite{DBLP:conf/nsdi/AgarwalDJSW10} & U & & & & & \checkmark    &&&\checkmark&&&\\\hline

\multicolumn{13}{|c|}{Scheduling for geo-distributed MapReduce-based systems (Section~\ref{subsec:Scheduling for Geo-distributed MapReduce-based Systems})}\\\hline

WANalytics~\cite{DBLP:conf/cidr/VulimiriCGKV15} & P \& U & & & & \checkmark & \checkmark    &&&&&&\checkmark \\\hline

Shuffle-aware data pushing~\cite{heintz2015end} & U & & & & & \checkmark    &&&&&&\\\hline

ViNe~\cite{DBLP:journals/internet/KeaheyTMF09} & P \& U & & \checkmark$^p$ & \checkmark & &     &&&&&&\\\hline

Reseal~\cite{DBLP:conf/ipps/KettimuthuASF16} & P & & & &  &     &&&&&&\\\hline

Rout~\cite{DBLP:conf/icdcs/JayalathE13} & P \& U & & & & \checkmark & \checkmark    &&&\checkmark&&&\\\hline

Meta-MapReduce~\cite{metamr_sss-15} & P \& U & & & & & \checkmark       &&\checkmark&\checkmark&&&\\\hline

Zhang el al.~\cite{DBLP:conf/IEEEcloud/ZhangLLZSMGA14} & P & & & & &\checkmark     &&&&&&\\\hline

\multicolumn{13}{|c|}{Scheduling for geo-distributed Spark-based systems (Section~\ref{subsec:Scheduling for Geo-distributed Spark-based Systems})}\\\hline

Pixida~\cite{DBLP:journals/pvldb/KloudasRPM15} & P & & & & \checkmark  &     &&&\checkmark&&&\\\hline

Flutter~\cite{DBLP:conf/infocom/HuLL16} & P & & & & \checkmark & \checkmark     &&&&&&\\\hline

Lazy optimal algorithm~\cite{DBLP:conf/hpdc/HeintzCS15} & P & & & &  & \checkmark    &&&\checkmark&&&\\\hline

\multicolumn{13}{|c|}{Resource allocation mechanisms for geo-distributed systems (Section~\ref{subsec:Resource Allocation Mechanism for Distributed Hadoop})}\\\hline

Awan~\cite{DBLP:conf/ic2e/JonathanCW16} & P & & & &  & \checkmark     &&&&&&\\\hline

Gadre et al.~\cite{DBLP:journals/sigmetrics/GadreRP11} & P \& U & & & & \checkmark & \checkmark    &&&&&&\\\hline

Ghit et al.~\cite{DBLP:conf/sc/GhitYE12} & P \& U & & & & & \checkmark    &&&&&&\\\hline

\multicolumn{13}{|p{15cm}|}{\textbf{Notations.} P: Pre-located geo-distributed big-data. U: User-located geo-distributed big-data. \checkmark$^p$: Systems provide only partial security, not a complete secure and private solution (\textit{e}.\textit{g}., G-Hadoop and ViNE allow authentication and end-to-end security, respectively, while SEMROD allows sensitive data security in the context of a hybrid cloud).}\\\hline


\end{tabular}
\egroup
\B
\caption{Summary of geo-distributed big-data processing frameworks and algorithms.}
\label{table:Summary}
\BBB\B
\end{center}
\end{table*}

\begin{description}[nolistsep,noitemsep,leftmargin=0.14in]
  \item[Pre-located geo-distributed big-data.] This category deals with data that is \textit{already} geo-distributed before the computation. For example, if there are $n$ locations, then all the $n$ locations have their data.
  \item[User-located geo-distributed big-data.] This category deals with frameworks that \textit{explicitly} distribute data to geo-locations before the computation begins. For example, if there are $n$ locations, then the user distributes data to the $n$ locations.
\end{description}
Note that there is a clear distinction between the above-mentioned two categories, as follows: The first category requires the distribution of jobs (\emph{not data}) over different clouds by the user site and then aggregation of outputs of all the sites at a specified site. The second category requires the distribution and/or partitioning of \emph{both} the data as well as jobs over different clouds by the user site. Here, an aggregation of outputs of all the sites is not must and depends on the job, if the job is partitioning the data over the clouds.

In the first category, we see MapReduce-based frameworks (\textit{e}.\textit{g}., G-Hadoop, GMR, Nebula, Medusa), Spark-based system (\textit{e}.\textit{g}., Iridium), a system for processing SQL-queries. As we mentioned, all these systems require to distribute a job over multiple clouds and then aggregation of outputs. In the second category, we see frameworks that do user-defined data and computation partitioning for achieving (\textit{i}) a higher level of fault-tolerance (by executing a job on multiple clouds, \textit{e}.\textit{g}., HMR, Spanner, and F1), (\textit{ii}) a secure computation by using public and private clouds (\textit{e}.\textit{g}., SEMROD), and (\textit{iii}) the lower job cost by accessing grid-resources in an opportunistic manner (\textit{e}.\textit{g}., HOG, KOALA grid-based system, and HybridMR).

A comparison of frameworks and algorithms for geo-distributed big-data processing based on several parameters such as security and privacy of data, data locality, selection of an optimal path for data transfer, and resource management is given in Table~\ref{table:Summary}.

\BB
\subsection{Frameworks for Pre-Located Geo-Distributed Big-Data}
\label{subsec:Frameworks for Pre Geo-Distributed Big-Data}

\subsubsection{Geo-distributed batch processing MapReduce-based systems for pre-located geo-distributed data}
\label{subsubsec:Implicit distributed data}
\noindent\textbf{G-Hadoop.} Wang et al. provided G-Hadoop~\cite{g-hadoop} framework for processing geo-distributed data across multiple cluster nodes, without changing existing cluster architectures. On the one hand, G-Hadoop processes data stored in a geo-distributed file system, known as Gfarm file system. On the other hand, G-Hadoop may increase fault-tolerance by executing an identical task in multiple clusters.

G-Hadoop consists of a G-Hadoop master node at a central location (for accessing G-Hadoop framework) and G-Hadoop slave nodes (for executing MapReduce jobs). The G-Hadoop master node accepts jobs from users, splits jobs into several sub-jobs and distributes them across slave nodes, and manages metadata of all files in the system. The G-Hadoop master node contains a metadata server and a global JobTracker, which is a modified version of Hadoop's original JobTracker. A G-Hadoop slave node contains a TaskTracker, a local Job Tracker, and an I/O server.

The Gfarm file system is a master-slave based distributed file system designed to share a vast amount of data among globally distributed clusters connected via a wide-area network. The master node called a Metadata Server (MDS) is responsible for managing the file system's metadata such as file names, locations, and access credentials. The MDS is also responsible for coordinating access to the files stored in the cluster. The multiple slave nodes, referred to as Data Nodes (DN), are responsible for storing raw data on local hard disks using local file systems. A DN runs a daemon that coordinates the access to the files on the local file system.

\smallskip\textit{Job execution in G-Hadoop}. Now, we discuss the job flow in G-Hadoop, which will help readers to understand a job execution in the geo-distributed environment. The job flow consists of three steps, as follows:
\begin{enumerate}[nolistsep,noitemsep,leftmargin=0.14in]
  \item \textit{Job submission and initialization}. The user submits a job to the G-Hadoop master node that creates a unique ID for the new job, and then, the user copies the map and reduce functions, job configuration files, and input files to a designated working directory at the master node of Gfarm file system. The global JobTracker initializes and splits the job.

  \item \textit{Sub-job assignment}. TaskTrackers of the G-Hadoop slaves request the global JobTracker for new tasks, periodically. The task assignment problem also considers the data locations. When a TaskTracker receives tasks, it copies executables and resources from the working directory of Gfarm file system.

  \item \textit{Sub-job execution}. Now, a computing node executes an assigned MapReduce task, as follows: (\textit{i}) a map task: it processes input data and writes outputs to a shared directory in the cluster; (\textit{ii}) reduce task: it contacts TaskTrackers that have executed the corresponding map tasks and fetches their outputs. If the TaskTracker is located in an identical cluster where data and reduce tasks are assigned, the data is read from the common shared directory of the cluster. Otherwise, the data is fetched using an HTTP request. The results of a reduce task are written to Gfarm file system.
\end{enumerate}

\noindent\emph{Pros}. G-Hadoop provides an efficient geo-distributed data processing, regards data locality, and hence, performs the map phase at the local site. G-Hadoop has a security mechanism, thereby an authenticated user can get access to only authorized data.

\noindent\emph{Cons}. G-Hadoop randomly places reducers in involved DCs~\cite{DBLP:journals/fgcs/ZhangZHJW16}. Also, it does not support iterative queries and HDFS, which is a common data storage, instead keeps the data in a new type of file system, Gfarm file system.

\begin{figure}[h]
\centering
\BB
\includegraphics[scale=0.4]{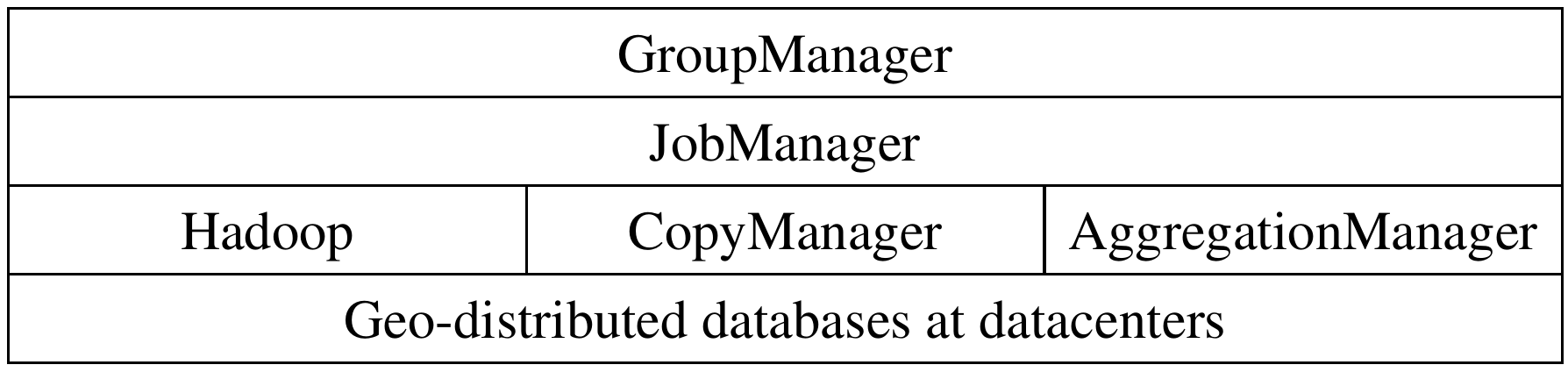}
\B
\caption{G-MR.}
\label{fig:gmr}
\BB
\end{figure}

\smallskip\noindent\textbf{G-MR.} G-MR~\cite{DBLP:journals/tc/JayalathSE14}, see Fig.~\ref{fig:gmr}, is a Hadoop-based framework that executes MapReduce jobs on a geo-distributed dataset across multiple DCs. Unlike G-Hadoop~\cite{g-hadoop}, G-MR does not place reducers randomly and uses a single directional weighted graph for data movement using the shortest path algorithm. G-MR deploys a GroupManager at a single DC and a JobManager at each DC. The GroupManager distributes the code of mappers and reducers to all the DCs and executes a data transformation graph (DTG) algorithm. Each JobManager manages and executes assigned local MapReduce jobs using a Hadoop cluster. Each JobManager has two components, namely a CopyManager for copying outputs of the job of one DC to other DCs and an AggregationManager for aggregating results from DCs.

The DTG algorithm finds an optimized path based on characteristics of the dataset, MapReduce jobs, and the DC infrastructure, for executing MapReduce jobs. The DTG algorithm constructs a graph by taking all the possible execution paths for executing the job. A node of the graph shows the number of MapReduce phases that have applied to input data and the data locations, and a weighted edge shows the computation flow. After constructing the graph, the problem of finding an optimized path for executing the job is reduced in finding a minimum weight path, which can be solved using the shortest path algorithm for the graph.

\textit{Execution steps}. A user submits G-MR codes to one of the DCs that executes the GroupManager. The GroupManager executes the DTG algorithm and determines the best path for collecting outputs from all the DCs. The GroupManager informs a JobManager of a DC about (\textit{i}) MapReduce jobs, (\textit{ii}) the local data that should be accessed by the job, and (\textit{iii}) where to copy the outputs of the job. The JobManagers execute the job accordingly using Hadoop, and then, use their local AggregationManager and CopyManager components for executing the respective tasks. Eventually, the GroupManager holds outputs of all the remaining DCs and performs the final computation to provide the final output.

\noindent\emph{Pros}. G-MR is a fully-functional and geo-distributed Hadoop-based framework.

\noindent\emph{Cons}. GMR is a non-secure framework and can only be used when the data is associative, \textit{i}.\textit{e}., the iterative and hierarchical reduce will not change the final result~\cite{DBLP:journals/fgcs/ZhangZHJW16}. Also, G-MR, like G-Hadoop, do not handle arbitrary and malicious faults, and cloud outages. These systems are unable to handle crash faults, similar to the standard MapReduce~\cite{DBLP:conf/ccgrid/Costa0RC16}.

\smallskip\noindent\textbf{Nebula.} Nebula~\cite{DBLP:conf/ic2e/RydenOCW14} is a system that selects the \textit{best node} for minimizing overall job completion time. Nebula consists of four centralized components, as follows: Nebula central, compute pool master, data-store master, and Nebula monitor. The Nebula central accepts jobs from the user who also provides the location of geo-distributed input data to the data-store master. The cmpute nodes, which are geo-distributed, periodically contact with the compute pool master, which is aware of all computing nodes in the system, and ask for jobs. A scheduler assigns tasks to the computing nodes based on the scheduling policy with the help of the compute pool master. Then, the computing nodes download the tasks and the input data from the data nodes according to specified locations by the data store master. When the computation finishes, the output is uploaded to data nodes, and the data-store master is informed of the location of the final outputs.

\smallskip\noindent\textbf{Medusa.} Medusa~\cite{DBLP:conf/ccgrid/Costa0RC16} system handles three new types of faults: processing corruption that leads to wrong outputs, malicious attacks, and cloud outages that may lead to the unavailability of MapReduce instances and their data. In order to handle such faults, a job is executed on $2f+1$ clouds, where $f$ faults are tolerable. In addition, a cloud is selected based on parameters such as available resources and bandwidth so that the job completion time is decreased.

\noindent\emph{Pros}. Medusa handles new types of faults.

\noindent\emph{Cons}. Except handling new types of faults, Medusa does not provide any new concept, and the fault handling systems can be included in a system that considers resource allocation and WAN traffic movement. The authors claim that they are not modifying the standard Hadoop; however, this claim is doubtful in the case of obtaining final outputs. The standard Hadoop system cannot produce the final outputs from partial outputs, \textit{e}.\textit{g}., equijoin of relations or finding maximum salaries of a person working in more than one department.

\subsubsection{Geo-distributed stream processing frameworks for pre-located geo-distributed data}
\label{subsubsec:A frameworks for implicit geo-distributed streaming data}
\noindent\textbf{Iridium.} Iridium~\cite{akella-15} is designed on the top of Apache Spark and consists of two managers, as follows: (\textit{i}) a \textit{global manager} is located in only one site for coordinating the query execution across sites, keeping track of data locations, and maintaining durability and consistency of data; and (\textit{ii}) a \textit{local manager} is located at each site for controlling local resources, periodically updating the global manager, and executing assigned jobs. Iridium considers heterogeneous bandwidths among different sites and optimizes data and task placement, which results in the minimal data transfer time among the sites. The task placement problem is described as a linear program that considers site bandwidths and query characteristics. An iterative greedy heuristic is used to move small chunks of datasets to sites having more bandwidth, resulting in efficient data transfer, without affecting the job completion time. Iridium speeds up processing by 64\% to 92\% as compared to Conviva~\cite{conviva}, Bing Edge, TPC-DS~\cite{tpc} and Berkeley Big Data Benchmark~\cite{bdb}, when deployed across eight Amazon Elastic Compute Cloud (EC2) regions in five continents. Iridium saves WAN bandwidth usage by 15\% to 64\%.

\noindent\emph{Pros}. While minimizing the job completion time, Iridium considers data and task placement regarding different bandwidth among DCs.

\noindent\emph{Cons}. Iridium considers the network congestion within a DC only, not among DCs. Also, Iridium minimizes only latency and does not consider the network bandwidth optimally~\cite{DBLP:journals/pvldb/KloudasRPM15,DBLP:conf/infocom/HuLL16}.

\smallskip The following frameworks, which are not based on Spark, are also designed for geo-distributed stream data processing where the data already exist at multiple locations.

\smallskip\noindent\textbf{JetStream.} JetStream~\cite{DBLP:conf/nsdi/RabkinASPF14} system processes geo-distributed streams and regards the network bandwidth and data quality. JetStream has three main components, as follows: geo-distributed workers, a centralized coordinator, and a client. The data is stored in a structured database of the form of a datacube. A client program creates a dataflow graph and submits it for the execution to the centralized coordinator. The coordinator selects linked-dataflow operators for each worker and then sends a relevant subset of the graph to each worker. Then, a worker creates necessary network connections with other workers and starts the operators. The execution terminates when the centralized coordinator sends a stop message or all the sources send a stop marker indicating that there will be no more data.

\noindent\emph{Pros}. JetStream, like Iridium, minimizes the amount of inter-DC traffic, but the approach is different from Iridium. JetStream uses data aggregation and adaptive filtering that support efficient OLAP queries as compared to Iridium.

\noindent\emph{Cons}. JetStream provides some degree of inaccuracy in the final results because of dropping and sampling results, hence, it is also good for small sensor networks. Unlike Iridium, JetStream does not support arbitrary SQL queries and does not optimize data and task placement~\cite{akella-15}. However, both, Iridium and JetStream do not deal with the network traffic and user-perceived delay simultaneously~\cite{DBLP:conf/cloud/HungGY15,DBLP:conf/hpdc/HeintzCS15}.


\smallskip\noindent\textbf{SAGE.} SAGE~\cite{DBLP:conf/ipps/TudoranAB13} is a \emph{general-purpose} cloud-based architecture for processing geo-distributed \textit{stream} data. SAGE consists of two types of services, as follows: (\textit{i}) \textit{Processing services} process incoming streaming data by applying the users' processing functions and provide outputs. Several queues at each geo-location handle stream data, where each processing service has one or more incoming queues. In addition, data is transformed into the required system format; and extract, transform, and load (ETL) software, \textit{e}.\textit{g}., IBM's InfoSphere DataStage~\cite{ibm-info}, are used for transforming data into the required format. (\textit{ii}) A \textit{global aggregator service} computes the final result by aggregating the outputs of the processing services. This process is executed in a DC nearby the user-location.

\noindent\emph{Pros}. Sage is independent of a data format, unlike JetStream, and also performs aggregation operation.

\noindent\emph{Cons}. Sage is designed to work with a limited number of DCs. The above-mentioned three stream processing frameworks perform data analytics over multiple geo-distributed sites, and the final computation is carried out at a single site. In the context of a large-scale IoT system, many sensors are widely distributed and send their expected results very often, \textit{e}.\textit{g}., location tracking systems. However, the current stream processing systems are not capable of handling streaming from such a huge number of devices~\cite{DBLP:conf/wf-iot/ChengPCK15,DBLP:journals/tetc/LiDOG16}.

\smallskip\noindent\textbf{G-cut.} G-cut~\cite{g-cut} proposed a way for allocating tasks in stream processing systems, specifically, for graph processing. Unlike Iridium that focuses on a general problem on a particular implementation (Spark), G-cut focuses on graph partitioning over multiple-DCs while minimizing inter-DC bandwidth usages and achieving user-defined WAN usages constraints. The algorithm consists of two phases: in the first phase, a stream graph processing algorithm does graph partitioning while satisfying the criteria of minimum inter-DC traffic and regarding heterogeneous bandwidth among DCs, and the second phase is used to refine the graph partitioning obtained in the first phase.

\subsubsection{SQL-style processing framework for pre-located geo-distributed data}
\label{subsubsec:A framework for implicit geo-distributed data processing using SQL}
\noindent\textbf{Geode.} Geode~\cite{DBLP:conf/nsdi/VulimiriCGJPV15} consists of three centralized components, as follows: a central command layer, pseudo-distributed measurement of data transfer, and a workload optimizer. The main component of Geode is the central command layer that receives SQL analytical queries from the user, partitions queries to create a distributed query execution plan, executes this plan over involving DCs, coordinates data transfers between DCs, and collates the final output. At each DC, the command layer interacts with a thin proxy layer that facilitates data transfers between DCs and manages a local cache of intermediate query results used for data transfer optimization. The workload optimizer estimates the current query plan or the data replication strategy against periodically obtained measurements from the command layer. These measurements are collected using the pseudo-distributed execution technique. Geode is built on top of Hive and uses less bandwidth than centralized analytics in a Microsoft production workload, TPC-CH~\cite{DBLP:conf/sigmod/ColeFGGKKKNNPSSSW11}, and Berkeley Big Data Benchmark~\cite{bdb}.

\noindent\emph{Pros}. Geode performs analytical queries locally at the data site. Also, Geode provides a caching mechanism for storing intermediate results and computing differences for avoiding redundant transfers. The caching mechanism reduces the data transfer for the given queries by 3.5 times.

\noindent\emph{Cons}. Geode does not focus on the job completion time and iterative machine learning workflows. 

%

\BB
\subsection{Frameworks for User-Located Geo-Distributed Big-Data}
\label{subsubsec:explicit distributed data}
In many cases, a single cluster is not able to process an entire dataset, and hence, the input data is partitioned over several clusters (possibly at different locations), having different configurations. In addition, geo-replication becomes necessary for achieving a higher level of fault-tolerance, because services of a DC may be disrupted for a while~\cite{DBLP:conf/sigmod/ZakharyNAA16,DBLP:journals/computer/Madria16}. In this section, we review frameworks that distribute data to geo-distributed clusters of different configurations, and hence, a user can select machines based on CPU speed, memory size, network bandwidth, and disk I/O speed from different locations. Geo-replication also ensures that a single DC will not be overloaded~\cite{DBLP:conf/nsdi/AgarwalDJSW10,DBLP:conf/sigmetrics/PotharajuJ13}. A system that distributes data to several locations must address the following questions at the time of design:
\begin{enumerate}[nolistsep,noitemsep,leftmargin=0.14in]
  \item How to store the input data, the intermediate data, and the final results?
  \item How to address shared data, data inter-dependencies, and application issues~\cite{DBLP:conf/nsdi/AgarwalDJSW10,CPE15}?
  \item How to schedule a task and where to place data? Answers to these questions impact job completion time and the cost significantly.
\item How to deal with task failures caused by using different clouds of non-identical configurations?
  \end{enumerate}
In addition, these systems must address inherent questions, \textit{i}.\textit{e}., how to efficiently aggregate the outputs of all the locations, how to deal with variable network bandwidth, and how to achieve strong consistency? Further details about geo-replication may be found in~\cite{ma}.

\subsubsection{Geo-distributed batch processing MapReduce-based systems for user-located geo-distributed data}
\label{subsubsec:Geo-distributed batch processing MapReduce-based systems for user located geo-distributed data}

\begin{figure}[h]
\centering
\BB
\includegraphics[scale=0.4]{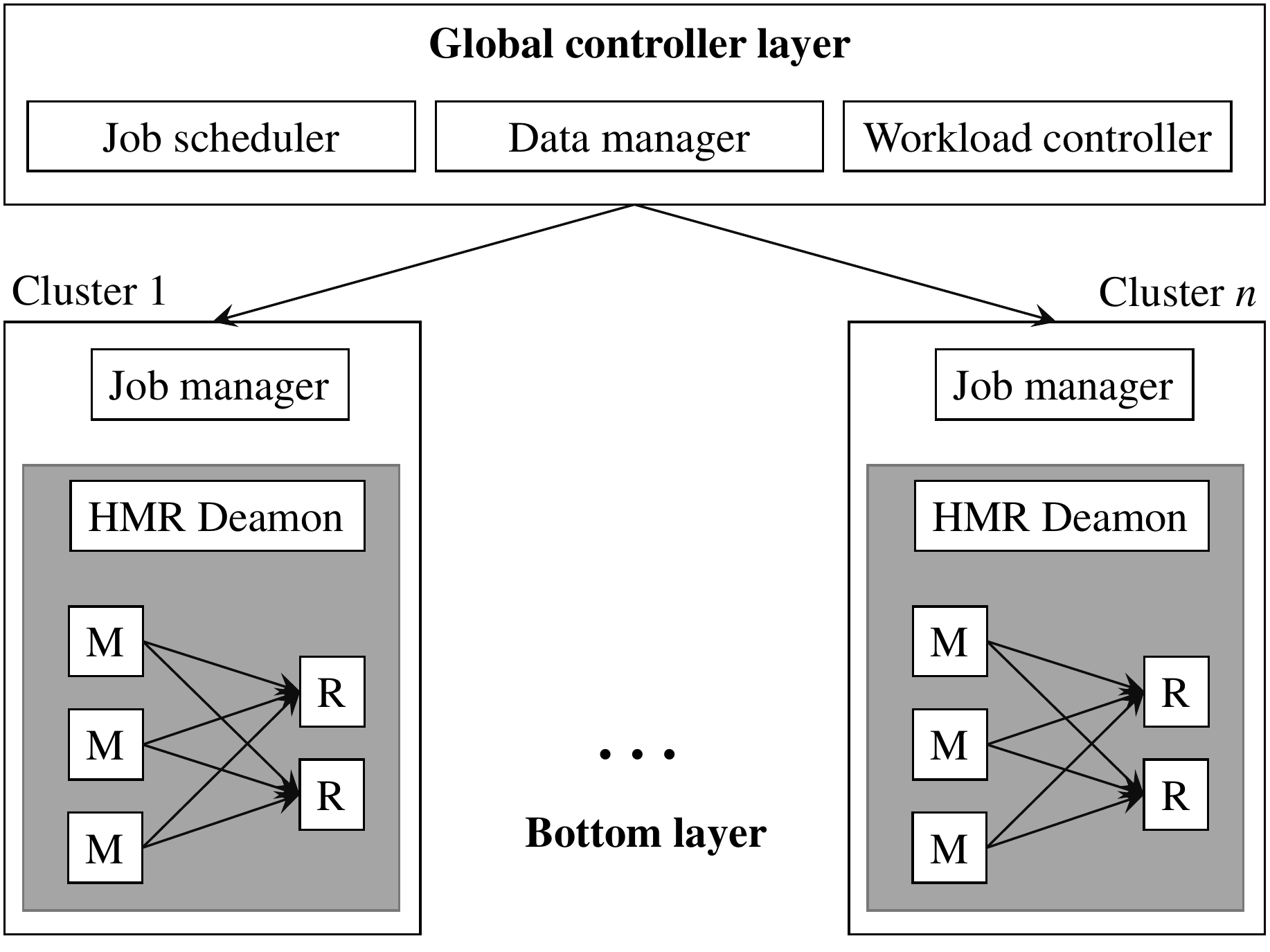}
\B
\caption{Hierarchical MapReduce programming model.}
\label{fig:hmr}
\BB
\end{figure}

\noindent\textbf{HMR.} Hierarchical MapReduce (HMR)~\cite{hmr} is a two-layered programming model, where the top layer is the global controller layer and the bottom layer consists of multiple clusters that execute a MapReduce job; see Fig.~\ref{fig:hmr}. A MapReduce job and data are submitted to the global controller, and the job is executed by the clusters of the bottom layer. Specifically, the global controller layer has three components, as follows: (\textit{i}) a job scheduler: partitions a MapReduce job and data into several sub-jobs and subsets of the dataset, respectively, and assigns each sub-job and a data subset to a local cluster; (\textit{ii}) a data manager: transfers map and reduce functions, job configuration files, and a data subset to local clusters; and (\textit{iii}) a workload controller: does load balancing. In the bottom layer, a job manager in a local cluster executes an HMR daemon and a local MapReduce sub-job. When the local sub-job is finished in a local cluster, the local cluster moves the final outputs to one of the local clusters that executes a global reducer for providing the final output.

\noindent\emph{Pros}. HMR is a trivial framework for geo-distributed MapReduce map-intensive jobs.

\noindent\emph{Cons}. HMR requires a full MapReduce job using the identity mapper to be executed before the global reducer. HMR is not efficient if intermediate data at different sites is huge and needs to be transferred to the global reducer, which is at a single site, resulting in the network bottleneck. In HMR, we also need to explicitly install a daemon on each one of the DCs.

\smallskip A simple extension to HMR is proposed in~\cite{DBLP:conf/iscc/CavalloMPT16}, where the authors suggested to consider the amount of data to be moved and the resources required to produce the final output at the global reducer. However, like HMR, this extension does not consider heterogeneous inter-DC bandwidth and available resources at the clusters. Another extension to both the systems is provided in~\cite{DBLP:conf/bdc/CavalloPMT16}, where the authors included clusters' resources and different network link capacity into consideration.

\begin{figure}[h]
\centering
\BB
\includegraphics[scale=0.45]{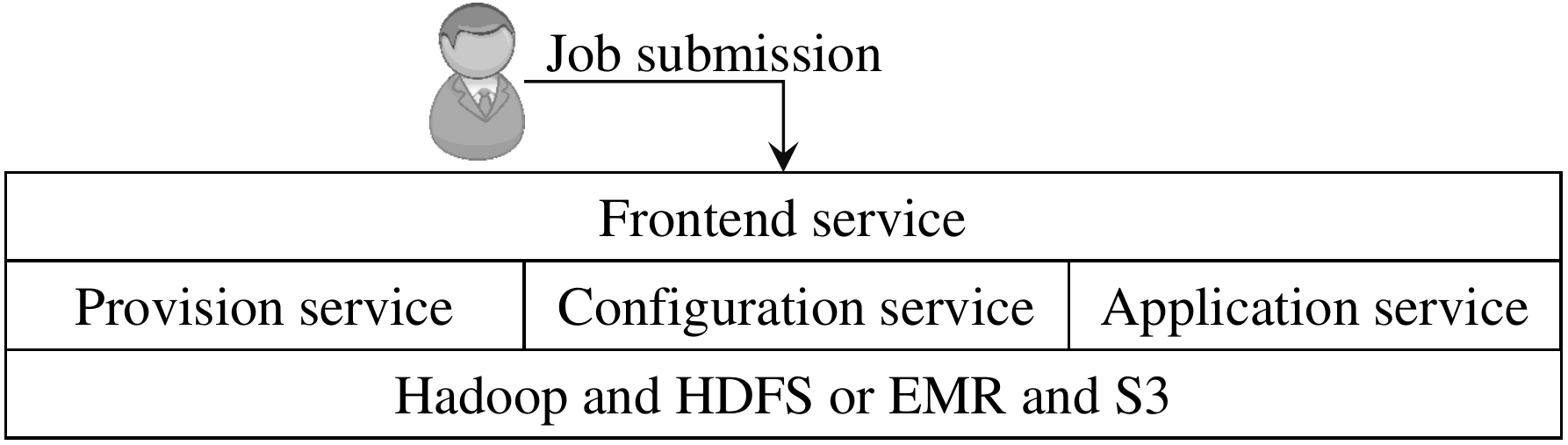}
\B
\caption{Resilin architecture.}
\label{fig:resilin}
\BB
\end{figure}

\smallskip\noindent\textbf{Resilin.} Resilin~\cite{resilin} provides a hybrid cloud-based MapReduce computation framework. Resilin; see Fig.~\ref{fig:resilin}, implements Amazon Elastic MapReduce (EMR)~\cite{amazon-emr} interface and uses the existing Amazon EMR tools for interacting with the system. In particular, Resilin allows a user to process data stored in a cloud with the help of other clouds' resources. In other words, Resilin partitions data as per the number of available clouds and moves data to those sites, which perform the computation and send the partial outputs to the source site. Resilin implements four services, as follows: (\textit{i}) a \textit{provision service} for starting or stopping virtual machines (VM) for Hadoop; (\textit{ii}) a \textit{configuration service} for configuring VMs; (\textit{iii}) an \textit{application service} for handling job flow; and (\textit{iv}) a \textit{frontend service} for implementing Amazon EMR API and processing users' requests.

\noindent\emph{Pros}. Resilin provides a way for exploiting the best available public resources. The major advantage of Resilin over EMR is that users can dynamically handle VMs as per needs, select different types of VMs, operating systems, and Hadoop versions.

\noindent\emph{Cons}. Resilin cannot be directly implemented to process geo-distributed data and requires further enhancements, which are not presented in~\cite{resilin}. Resilin does not provide data security by dealing with data sensitivity that can be explored in a hybrid setting. The next hybrid cloud based system handles data sensitivity and provides a secure solution.

\smallskip\noindent\textbf{SEMROD.} SEMROD~\cite{DBLP:conf/sigmod/OktayMKK15} first finds sensitive and non-sensitive data and sends non-sensitive data to public clouds. Private and public clouds execute the map phase. In order to hide, some keys that are required by the private cloud, the public cloud sends all the outputs of the map phase to the private cloud only in the first iteration. The private cloud executes the reduce phase only on sensitive key records and ignores non-sensitive keys. For example, let $k_1$ and $k_2$ are two keys at the public cloud, and $k_2$ also exists at the private cloud. The public cloud will send $\langle \mathit{key}, \mathit{value}\rangle$ pairs of $k_1$ and $k_2$ to the private cloud that will perform the reduce phase only on $k_1$. Public clouds, also, execute the reduce phase on all the outputs of the map phase. At the end, a filtering step removes duplicate entries, created by the transmission of the public mappers' outputs.

\noindent\emph{Pros}. By storing sensitive data in the private cloud, SEMROD provides a secure execution and performs efficiently if the non-sensitive data is smaller than the sensitive data. Note that SEMROD is not the first hybrid cloud solution for MapReduce computations based on data sensitivity. HybrEx~\cite{DBLP:conf/hotcloud/KoJM11}, Sedic~\cite{DBLP:conf/ccs/ZhangZCWR11}, and Tagged-MapReduce~\cite{DBLP:conf/ccgrid/ZhangCY14} are also based on data sensitivity. However, they are not secure because during the computation they may leak information by transmitting \emph{some} non-sensitive data between the private and the public cloud, and this is the reason we do not include HybrEx, Sedic, and Tagged-MapReduce in this survey.

\noindent\emph{Cons}. The transmission of the whole outputs of the map phase to the private cloud is the main drawback of SEMROD. If only a few keys are required at the private cloud, then it is useless to send entire public side outputs to the private cloud.

\smallskip\noindent\textbf{HOG.} Hadoop on the Grid (HOG)~\cite{hog} is a geo-distributed and dynamic Hadoop framework on the grid. HOG accesses the grid's resources in an opportunistic manner, \textit{i}.\textit{e}., if users do not own resources, then they can opportunistically execute their jobs, which can be preempted at any time when the resource owner wants to execute a job. HOG is executed on the top of Open Science Grid (OSG)~\cite{osg}, which spans over 109 sites in the United States and consists of approximately 60,000 CPU cores. HOG has the following components:
\begin{enumerate}[nolistsep,noitemsep,leftmargin=0.14in]
  \item \textit{Grid submission and execution component}. This component handles users' requests, allocation and deallocation of the nodes on the grid, which is done by transferring a small-sized Hadoop executables package, and the execution of a MapReduce job. Further, this component dynamically adds or deletes nodes according to an assigned job requirement.

  \item \textit{HDFS}. HDFS is deployed across the grid. Also, due to preemption of tasks, which results in a higher node failure, the replication factor is set to 10. A user submits data to a dedicated node using the grid submission component that distributes the data in the grid.

  \item \textit{MapReduce-based framework}. This component executes MapReduce computations across the grid.
\end{enumerate}
\noindent\emph{Pros}. HOG (and the following two grid-based systems) consider that a single DC is distributed across multiple DCs while the previously reviewed frameworks support multiple DCs collaborating for a task.

\noindent\emph{Cons}. The multi-cluster HOG processing is only for fault-tolerance; however, there is \emph{no} real multi-cluster processing in HOG so that no parallel processing, and hence, no necessity for aggregating the site outputs.

\begin{figure}[!t]
\centering
\BB
\includegraphics[scale=0.4]{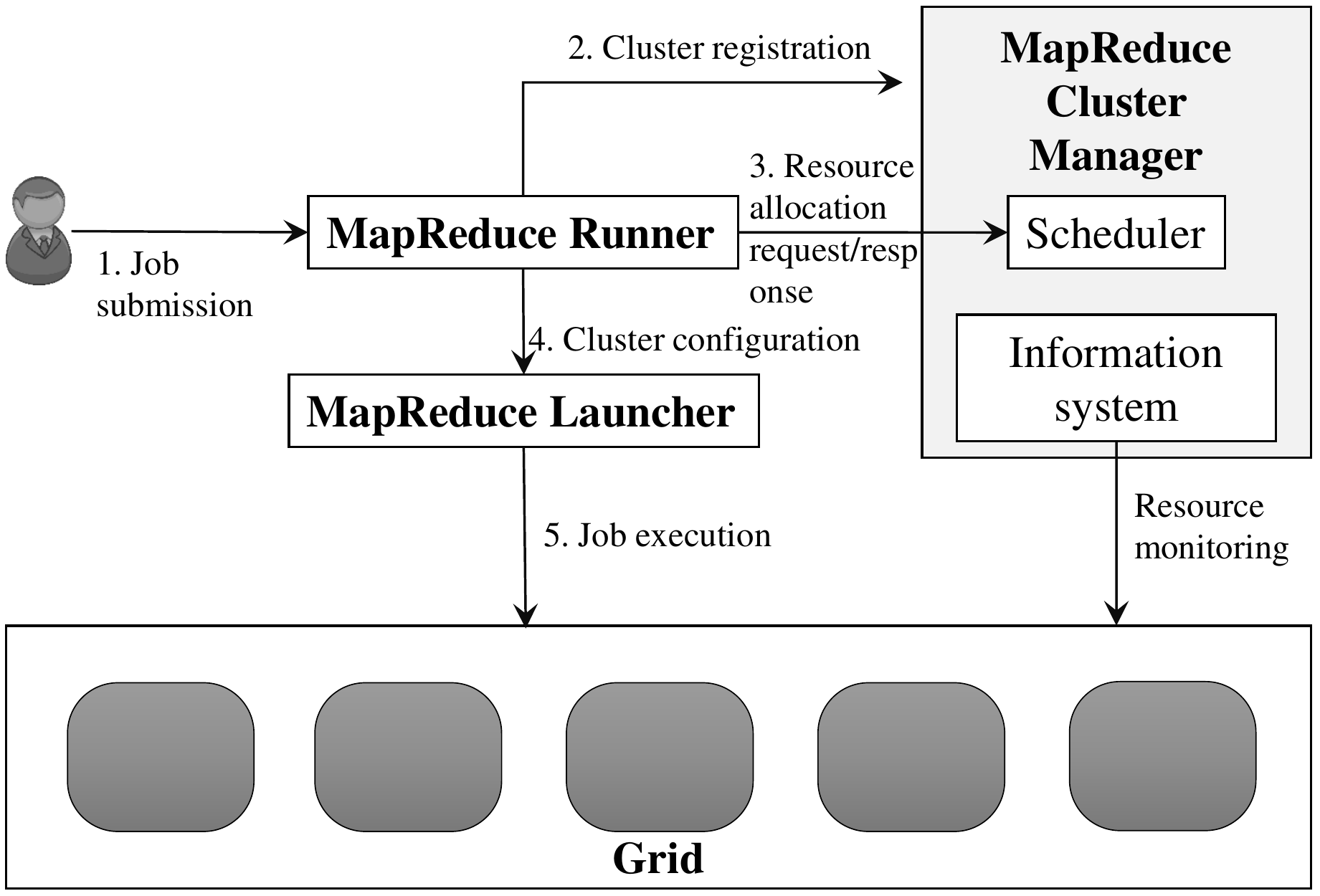}
\B
\caption{A multi-cluster MapReduce architecture based on KOALA grid scheduler.}
\label{fig:koala}
\BB
\end{figure}

\smallskip\noindent\textbf{KOALA grid-based system.} Ghit et al.~\cite{DBLP:conf/sc/GhitYE12} provided a way to execute a MapReduce computation on KOALA grid~\cite{DBLP:journals/concurrency/MohamedE08}. The system has three components, as shown in Fig.~\ref{fig:koala}, MapReduce-Runner, MapReduce-Launcher, and MapReduce-Cluster-Manager.

\noindent\textit{MapReduce-Runner}. MapReduce-Runner interacts with a user, KOALA resource manager, and the grid's physical resources via MapReduce-Launcher. It deploys a MapReduce cluster on the grid with the help of MapReduce-Launcher and monitors parameters such as the total number of (real) MapReduce jobs, the status of each job, and the total number of map and reduce tasks. MapReduce-Runner designates one node as the master node and all the other nodes as slave nodes.

\noindent\textit{MapReduce-Launcher}. MapReduce-Launcher is responsible for configuring a distributed file system on the grid resources and a compute framework. In addition, MapReduce-Launcher executes an assigned job on the grid and turns off the cluster after the execution.

\noindent\textit{MapReduce-Cluster-Manager}. MapReduce-Cluster-Manager is a central entity, which stays in the scheduler site for maintaining the grid, metadata of each MapReduce cluster. MapReduce-Cluster-Manager is also responsible for growing or shrinking the nodes in a cluster, with the help of the KOALA Information System module.

\noindent\emph{Pros}. KOALA grid-based system provides scheduling of multiple jobs as a part of single MapReduce instances~\cite{DBLP:conf/jsspp/KuzmanovskaME14}. Also, it provides a way for performance, data, failure and version isolation in the grid settings.

\noindent\emph{Cons}. KOALA grid-based system and the previous systems, HOG, are dependent on a special grid architecture. Also, the practicality of these systems in the context of public clouds is not known.

\smallskip\noindent\textbf{HybridMR.} HybridMR~\cite{CPE15} allows a MapReduce job on a desktop grid and cloud infrastructures simultaneously. HybridMR consists of two layers, as follows: (\textit{i}) a \textit{service layer} that allows data scheduling, job scheduling, metadata storage, and database storage to a new distributed file system, called HybridDFS; (\textit{ii}) a \textit{resource layer} that contains reliable cluster nodes and many unreliable volunteer/desktop nodes. A user uploads MapReduce jobs and data into HybridMR. The data scheduler assigns data to cloud nodes and desktop nodes on which jobs are scheduled by the job scheduler.

\noindent\emph{Pros}. Unlike KOALA grid-based system and HOG, HybridMR provides a way to execute a job on the cloud as well as on the grid.

\noindent\emph{Cons}. Resilin, KOALA grid-based system, HOG, and HybridMR execute a MapReduce job using a modified version of the standard Hadoop on grid systems in an opportunistic manner, which a challenging task, because resource seizing may delay the entire job. In addition, all the grid-based systems, such as mentioned above, suffer from inherited limitations of a grid, \textit{e}.\textit{g}., accounting and administration of the grid, security, pricing, and a prior knowledge of resources. Moreover, the grid-based systems include new nodes during computations. However, adding nodes without data locality during the execution may reduce the job performance, resulting in no gain from inter-DC scaling~\cite{guo2015moving}.

\subsubsection{Geo-distributed stream processing frameworks for user-located geo-distributed data}
\label{subsubsec:Geo-distributed stream processing frameworks for user-located geo-distributed data}
In order to give a flavor of stream processing in user-located geo-distributed data, we include Photon~\cite{DBLP:conf/sigmod/AnanthanarayananBDGJQRRSV13} that is not built on top of Apache Spark.

\noindent\textbf{Google's Photon.} Google's Photon~\cite{DBLP:conf/sigmod/AnanthanarayananBDGJQRRSV13} is a highly scalable and very low latency system, helping Google Advertising System. Photon works on exactly-once semantics (that is only one joined tuple is produced) and handles automatic DC-level fault-tolerance.

Photon performs equijoin between a primary table (namely, the \textit{query event} that contains \texttt{query id}, \texttt{ads id}, and \texttt{ads text}) and a foreign table (namely, the \textit{click event} that contains \texttt{click id}, \texttt{query id}, and \texttt{user clicked log information}). Both the tables are copied to multiple DCs. Any existing streaming-based equijoin algorithm cannot join these two tables, because a click can only be joined if the corresponding query is available. In reality, the query needs to occur before the corresponding click, and that fact is not always true in the practical settings with Google, because the servers generating clicks and queries are not located at a single DC.

An identical Photon pipeline is deployed in multiple DCs, and that works independently without directly communicating with other pipelines. Each pipeline processes all the clicks present in the closest DCs and tries to join the clicks with the query based on the \texttt{query id}. Each pipeline keeps retrying until the click and query are joined and written to an \emph{IdRegistry}, which guarantees that each output tuple is produced exactly once.

Google's Mesa~\cite{DBLP:journals/pvldb/GuptaYGKCLWDKABHCSJSGVA14}, Facebook's Puma, Swift,
and Stylus~\cite{DBLP:conf/sigmod/ChenWIJLSWWWY16} are other industry deployed stream processing distributed frameworks. A brief survey of general approaches for building high availability stream processing systems with challenges and solutions (Photon, F1, and Mesa) is presented in~\cite{gupta2015high}.

\subsubsection{SQL-style processing framework for user-located geo-distributed data}
\label{subsubsec:SQL-style processing framework for user-located geo-distributed data}

\noindent\textbf{Google's Spanner.} Google's Spanner~\cite{DBLP:conf/osdi/CorbettDEFFFGGHHHKKLLMMNQRRSSTWW12} is a globally-distributed data management system. In~\cite{DBLP:conf/sigmod/BaconBBCDFFGJKL17}, database aspects, \textit{e}.\textit{g}., distributed query execution in the presence of sharding/resharding, query restarts upon transient failures, and range/index extraction, of Spanner are discussed.

In Spanner, \emph{table interleaving} is used to keep tables in the database, \textit{i}.\textit{e}., rows of two tables that will join based on a joining attribute are kept co-located, and then, tables are partitioned based on the key. Each partition is called a \emph{shard} that is replicated to multiple locations.

A new type of operation is introduced, called \emph{Distributed Union} that fetches results from all the shard according to a query. However, performing the distributed union before executing any other operations, \textit{e}.\textit{g}., scan, filter, group by, join, and top-k, will cause to read multiple shards, which may not participate in the final output. Hence, all such operators are pushed to the table before the distributed union, which takes place at the end to provide the final answer. Three different mechanisms of index or range retrieval are given, as follows: distribution range extraction, seek range extraction, and lock range extraction.

\smallskip
A recent paper~\cite{sharad-icde-17} carries the same flavor of the hybrid cloud computation, discussed in \S\ref{subsubsec:Geo-distributed batch processing MapReduce-based systems for user located geo-distributed data}, and suggests a general framework for executing SQL queries, specifically, select, project, join, aggregation, maximum, and minimum, while not revealing any sensitive data to the public cloud during the computation.

\BB
\section{Data Transfer Systems/Algorithms}
\label{subsec:Data Transfer Systems}
We reviewed G-MR (\S\ref{subsubsec:Implicit distributed data}) that finds the best way only for inter-cloud data transfer, but not based on real-time parameters. Tudoran et al.~\cite{DBLP:conf/ccgrid/TudoranCWBA14} and Gadre et al.~\cite{DBLP:conf/cac/GadreRMP13} proposed a data management framework for efficient data transfer among the clouds, where each cloud holds monitoring, data transfers, and decision management agents. The monitoring agent monitors the cloud environment such as available bandwidth, throughput, CPU load, I/O speed, and memory consumption. The decision agent receives the monitored parameters and generates a real-time status of the cloud network and resources. Based on the status, the decision agent finds a directed/multi-hop path for data transfer from the source to the destination. The transfer agent performs the data transfers and exploits the network parallelism. ViNe~\cite{DBLP:journals/internet/KeaheyTMF09} is the only system that offers end-to-end secure connectivity among the clusters executing MapReduce jobs.

\smallskip\noindent\textbf{Volley.} Volley~\cite{DBLP:conf/nsdi/AgarwalDJSW10} is a 3-phase iterative algorithm that places data across geo-distributed DCs. A cloud submits its logs to Volley that analyzes the logs using SCOPE~\cite{DBLP:journals/vldb/ZhouBWLCS12}, a scalable MapReduce-based platform, for efficient data transfer. Volley also includes real-time parameters, such as capacity and cost of all the DCs, latency among DCs, and the current data item location. The current data item location helps in identifying whether the data item requires movement or not. In phase 1, data is placed according to users' IP addresses at locations as closest as possible to the user. However, the data locations as a consequence of phase 1 are not the best in terms of closeness to the actual user's location. Hence, in phase 2, data is moved to the closest and best locations to the user via a MapReduce-based computation. Phase 3 is used to satisfy the DC capacity constraint, and if a DC is overloaded, then some data items that are not frequently accessed by the user are moved to another closest DC.

\noindent\emph{Pros}. Volley can be used in optimizing automatic data placement before a computation execution using any above-mentioned frameworks in \S\ref{subsubsec:explicit distributed data}.

\noindent\emph{Cons}. Volley does not consider bandwidth usage~\cite{DBLP:conf/sigcomm/BodikMCMMS12}, unlike JetStream~\cite{DBLP:conf/nsdi/RabkinASPF14}.

\begin{figure}[!h]
\centering
\B\B
\includegraphics[scale=0.4]{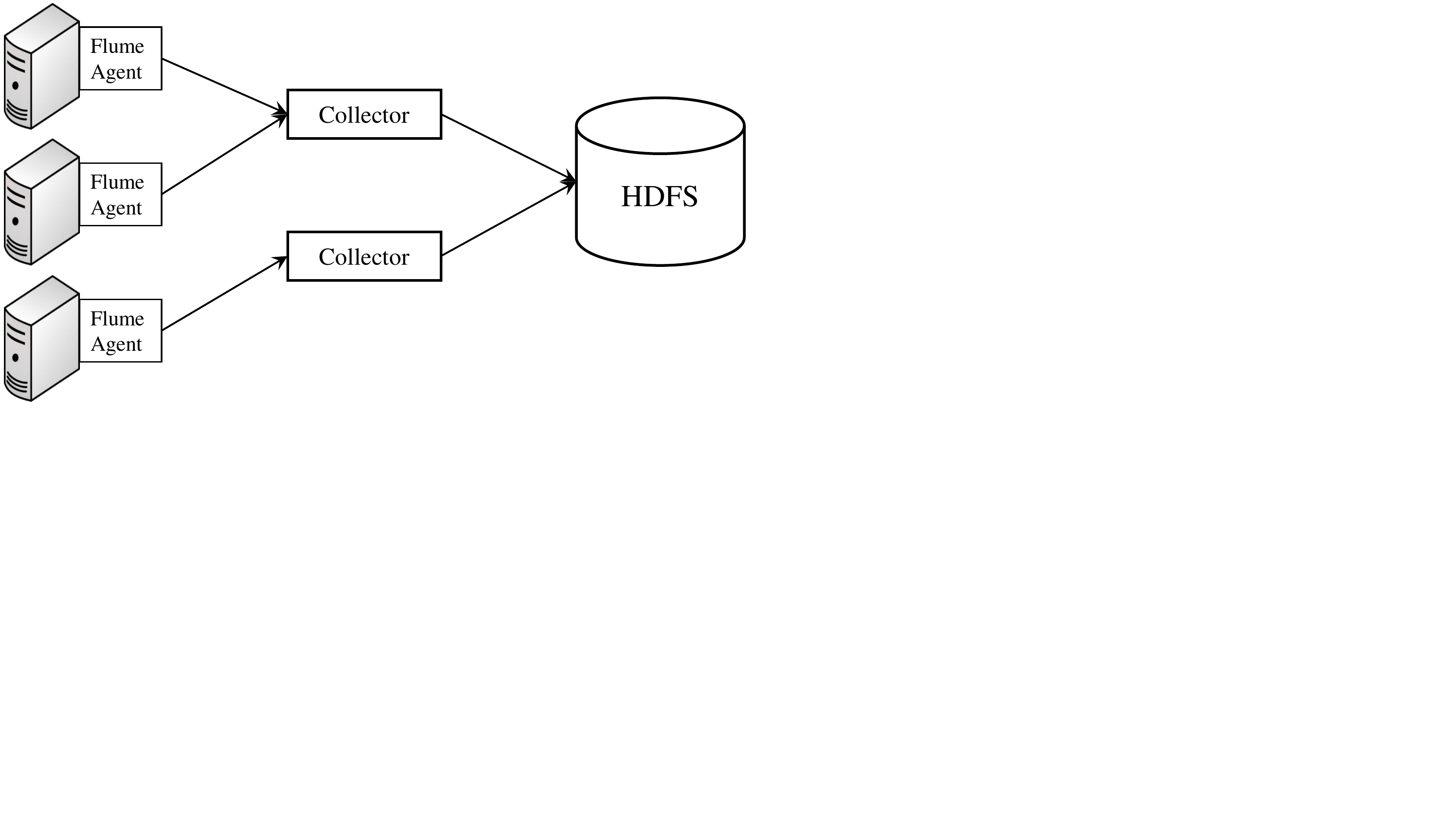}
\caption{Apache Flume.}
\label{fig:flume}
\BBB\B
\end{figure}

\smallskip\noindent\textbf{Apache Flume.} Apache Flume~\cite{apache_flume} is a distributed, reliable, scalable and available service for efficiently collecting, aggregating, and moving a large amount of log data from various sources to a centralized data store --- HDFS or HBase. However, it should be noted down that Flume is \emph{a general-purpose data collection service}, which can be used in geo-distributed settings. An \emph{event} is the basic unit of the data transported inside Flume. Events and log data are generated at different log servers that have Flume \emph{agents}, see Fig.~\ref{fig:flume}. Flume agents transfer data to intermediate nodes, called \emph{collectors}. The collector aggregates data and pushes this data to a centralized data store. Flume provides guaranteed data delivery and stores data in a buffer when the rate of incoming data exceeds the rate at which data can be written to the destination~\cite{apache_flume1}.

\BB
\section{Scheduling in Geo-Distributed Systems}
\label{subsec:Preprocessing-based Architecture}
In this section, we present some methods/architectures that preprocess a job before deploying it over distributed locations to find the best way for data distribution and/or the best node for the computation. The main idea of the following methods is in reducing the total amount of data transfer among DCs. Note that the following methods work offline and do not provide a way for executing a geo-distributed job on top of Hadoop/Spark, unlike systems in \S\ref{subsubsec:explicit distributed data} that execute a job and may handle such offline tasks too.

\BB
\subsection{Scheduling for Geo-distributed MapReduce-based Systems}
\label{subsec:Scheduling for Geo-distributed MapReduce-based Systems}

\noindent\textbf{WANalytics.} WANalytics~\cite{DBLP:conf/cidr/VulimiriCGKV15} preprocesses a MapReduce job before its real implementation and consists of two main components, as follows:

\noindent
\emph{Runtime analyzer}: executes user's job directed acyclic graph (DAG), which is about job execution flow, in a distributed way across DCs. The runtime analyzer finds a physical plan that specifies where does each stage of the job to be executed and how will data be transferred across DCs. The runtime layer consists of a centralized coordinator, only with one DC that interacts with all the other DCs. Users submit a DAG of jobs to the coordinator that asks the workload analyzer to provide a physical distributed execution plan for the DAG.

  \noindent
\emph{Workload analyzer}: continuously monitors and optimizes the user's DAG and finds a distributed physical plan according to the DAG. The plan is determined in a manner that minimizes the total bandwidth usage by considering DC locations and data replication factor.
\noindent\emph{Cons}. Unlike Iridium, WANalytics does not consider the network bandwidth and job latency, and only focuses on the amount of data transfer among DCs. In addition, WANalytics is not designed to handle iterative machine learning workflows~\cite{DBLP:journals/corr/CanoWMCF16}.

\smallskip\noindent\textbf{Shuffle-aware data pushing.} Heintz et al.~\cite{heintz2015end} suggested shuffle-aware data pushing at the map phase. It finds all those mappers that affect the job completion in a DC, and hence, rejects those mappers for a new job. In other words, the algorithm selects only mappers that can execute a job and shuffle the intermediate data under a time constraint. Mappers are selected based on monitoring the most recent jobs. The algorithm is presented for a single DC and can be extended to geo-distributed settings.

\noindent\emph{Cons}. It is assumed that the same mappers have appeared in previous jobs; otherwise, it is hard to have a prior knowledge of mappers.

\smallskip\noindent\textbf{Reseal.} Reseal~\cite{DBLP:conf/ipps/KettimuthuASF16} considers a bi-objective scheduling problem for scheduling response-critical (RC) tasks and best effort (BE) tasks. Each task is associated with a utility function that provides a value, which is a function of the task's slow down. A task's value is initially set to be high, and then, decreases over time if the task is delayed. Two approaches are suggested for allocating RC tasks, as follows: (\textit{i}) \textit{Instant-RC}: refers to the scheduling of a RC task over many BE tasks on the arrival the RC task. In other words, a RC task is allocated to have an identical throughput as it would achieve in the absence of any BE task in the system; (\textit{ii}) \textit{Threshold-RC}: refers to the scheduling of a RC task according to its utility function. In other words, a RC task is not allocated on its arrival, but scheduled in a manner that it finishes with a slowdown according to its utility function.

\noindent\emph{Pros}. Reseal is the first approach for dealing with realtime scheduling and regarding the network bandwidth in terms of BE transfers in the context of geo-distributed computations.

\smallskip\noindent\textbf{Error-bound vs staleness-bound algorithms.} In~\cite{DBLP:conf/cloud/HeintzCS16}, the authors considered the problem of adaptive data movement satisfying the timeliness vs accuracy under bandwidth limitations and presented two online algorithms: error-bound and staleness-bound algorithms. These algorithms are based on 2-levels of caching. The error-bound algorithm allows the insertion of new values to the second-level cache, and the values from the second-level cache are moved to the first-level cache when their aggregate values exceed the error constraint. In contrast, the staleness-bound algorithm dynamically finds the ranking of the second-level cache values by their (estimated) initial prefix error, and then, defines the first-level cache to comprise the top values from the second-level cache. Both the cache-based algorithms does not answer the following questions: (\textit{i}) how to define the size of the cache, is it application dependent or not? (\textit{ii}) how do the cache-based algorithms handle a huge amount of streaming data in an IoT environment, do the algorithms sustain any progressive computation on data or not.

\smallskip\noindent\textbf{Rout.} Jayalath and Eugster~\cite{DBLP:conf/icdcs/JayalathE13} extended Pig Latin~\cite{DBLP:conf/sigmod/OlstonRSKT08}, called Rout, by introducing geo-distributed data structures and geo-distributed operations. The authors suggest that before executing a geo-distributed job, it is beneficial to analyze the job, thereby the data transfer among DCs is reduced. A Rout program maximizes job parallelization by generating a set of MapReduce jobs and determines optimal points in the execution for performing inter-DC copy operations. A Rout program generates a MapReduce dataflow graph, like Pig Latin, and analyzes it for finding points, \textit{i}.\textit{e}., which DCs will perform inter-DC data transfer and to where. Based on the points, the dataflow graph is annotated, and then, an execution plan is generated to consider dynamic runtime information and transfer data to not overloaded DCs.

\noindent\emph{Pros}. Rout reduces the job completion time down to half, when compared to a straightforward schedule.

\smallskip\noindent\textbf{Meta-MapReduce.} Meta-MapReduce~\cite{metamr_sss-15} reduces the amount of data required to transfer between different locations, by transferring essential data for obtaining the result. Meta-MapReduce regards the locality of data and mappers-reducers and avoids the movement of data that does not participate in the final output. Particularly, Meta-MapReduce provides an algorithmic way for computing the desired output using metadata (which is exponentially smaller than the original input data) and avoids uploading the whole data. Thus, Meta-MapReduce enhances the standard MapReduce and can be implemented into the state-of-the-art MapReduce systems, such as Spark, Pregel~\cite{DBLP:conf/sigmod/MalewiczABDHLC10}, or modern Hadoop.

\noindent\emph{Pros}. A MapReduce job can be enhanced by sampling local data, which cannot be used for future analysis. However, designing good sampling algorithms is hard. Meta-MapReduce does not need any sampling, and hence, has a wide applicability.

\smallskip
Zhang el al.~\cite{DBLP:conf/IEEEcloud/ZhangLLZSMGA14} provided prediction-based MapReduce job localization and task scheduling approaches. The authors perform a sub-cluster-aware scheduling of jobs and tasks. The sub-cluster-aware scheduling finds sub-clusters that can finish a MapReduce job efficiently. The decision is based on several parameters such as the execution time of the map phase, the execution time of a DC remote map task, percentage of remote input data, number of map tasks in the job, and number of map slots in a sub-cluster. 

Li et al.~\cite{pli-2017} provided an algorithm for minimizing the shuffle phase inter-DC traffic by considering both data and task allocation problems in the context of MapReduce. The algorithm finds DCs having higher output to input ratio and poor network bandwidth, and hence, move their data to a \emph{good} DC. Note that the difference between this algorithm and Iridium~\cite{akella-15} is in considering an underlying framework. Chen et al.~\cite{DBLP:journals/tc/ChenPL16} also provided a similar algorithm and showed that the data local computations are not always best in a geo-distributed MapReduce job.

\BB
\subsection{Scheduling for Geo-distributed Spark-based Systems}
\label{subsec:Scheduling for Geo-distributed Spark-based Systems}

\noindent\textbf{Pixida.} Pixida~\cite{DBLP:journals/pvldb/KloudasRPM15} is a scheduler that minimizes data movement across resource constrained inter-DC links. Silos are introduced as the main topology. Silo considers each node of a single location as a super-node in a task-level graph. The edges between the super-nodes show the bandwidth between them. Hence, Pixida considers that sending data to a node within an identical silo is preferable than sending data to nodes in remote silos. Further, a variation of the min-k cut problem is used to assign tasks in a silo graph.

\smallskip\noindent\textbf{Flutter.} The authors suggested a scheduling algorithm, Flutter~\cite{DBLP:conf/infocom/HuLL16}, for MapReduce and Spark. This algorithm is network-aware and finds on-the-fly job completion time based on available compute resources, inter-DC bandwidth, and the amount of data in different DCs. At the time of the final computation assignment, Flutter finds a DC that results in the least amount of data transfer and having most of the inputs that participate in the final output.

\smallskip\noindent\textbf{Lazy optimal algorithm.} Lazy optimal algorithm~\cite{DBLP:conf/hpdc/HeintzCS15} considers a tradeoff between
the amount of inter-DCs data and staleness. Lazy optimal algorithm is based on two algorithms, as follows: (\textit{i}) traffic optimality algorithm: transfers exactly one update to the final computational site for each distinct key that arrived in a specified time window, and (\textit{ii}) eager optimal algorithm: transfers exactly one update for each distinct key immediately after the last arrival for that key within a specified time window. The lazy optimal algorithm makes a balance between the two algorithms and transfers updates at the last possible time that would still provide the optimal value of staleness. As a major advantage, the lazy optimal algorithm considers several factors such as the network bandwidth usage, data aggregation, query execution, and response
latency, and extends Apache Storm~\cite{Storm} for supporting efficient geo-distributed stream analytics~\cite{DBLP:conf/wf-iot/ChengPCK15}.

\BB
\subsection{Resource Allocation Mechanisms for Geo-Distributed Systems}
\label{subsec:Resource Allocation Mechanism for Distributed Hadoop}
\noindent\textbf{Awan.} Awan~\cite{DBLP:conf/ic2e/JonathanCW16} provides a resource lease abstraction for allocating resources to individual frameworks. In other words, Awan is a system that does not consider underlying big-data processing frameworks when allocating resources. Awan consists of four centralized components, as follows: (\textit{i}) file master, (\textit{ii}) node monitor, (\textit{iii}) resource manager, which provides the states of all resources for different frameworks, and (\textit{iv}) framework scheduler, which acquires available resources using a resource lease mechanism. The resource lease mechanism provides a lease time to each resource in which resources are only used by the framework scheduler during the lease only, and after the lease time, the resource must be vacated by the framework scheduler.

\smallskip Ghit et al.~\cite{DBLP:conf/sc/GhitYE12} provided three policies for dynamically resizing a distributed MapReduce cluster. As advantages, these policies result in less reconfiguration costs and handle data distribution in reliable and fault-tolerant manners.
\begin{itemize}[nolistsep,noitemsep,leftmargin=0.14in]
  \item \textit{Grow-Shrink Policy}. It is a very simple policy that maintains a ratio of the number of running tasks (map and reduce tasks) and the number of available slots (map and reduce slots). Based on the ratio, the system adds (or removes) nodes to (or from) the cluster.

  \item \textit{Greedy-Grow Policy}. This policy suggests adding a node to a cluster in a greedy manner. However, all the resources are added regardless of the cluster utilization.

  \item \textit{Greedy-Grow-with-Data Policy}. This policy adds core nodes, unlike the previous policy that adds only transient nodes. Hence, on resource availability, the node is configured for executing TaskTracker. However, the policy does not consider cluster shrink requests.
\end{itemize}

Ghit et al.~\cite{DBLP:conf/sigmetrics/GhitYIE14} extended the above-mentioned policies by accounting dynamic demand (job, data, and task), dynamic usage (processor, disk, and memory), and actual performance (job slowdown, job throughput, and task throughput) analysis when resizing a MapReduce cluster.

\smallskip Gadre et al.~\cite{DBLP:journals/sigmetrics/GadreRP11} provided an algorithm for assigning the global reduce task, thereby the data transfer is minimal. The algorithm finds the answer to the questions such as when to start the reduce phase for a job, where to schedule the global reduce task, which DC is holding a major part of partial outputs that participate in the final output, and how much time is required to copy outputs of DCs to a single (or multiple) location for providing the final outputs? During a MapReduce job execution, one of the DCs (working as a master DC) monitors all the remaining DCs and keeps the total size of outputs in each DC. Monitoring helps in identifying the most prominent DC while scheduling the global reduce phase.

\smallskip
\noindent\emph{Cons}. The Awan and the above-mentioned three policies do not answer a question: what will happen to a job if resources are taken during the execution? Also, these mechanisms do not provide a way for end-to-end overall
improvement of the MapReduce dataflow, load balancing, and cost-efficient data movement~\cite{saeed2016sandooq}. Gadre et al.~\cite{DBLP:journals/sigmetrics/GadreRP11} optimizes the reduce data placement according to map's output location, which might slow down the job due to the low bandwidth~\cite{DBLP:conf/ic2e/HeintzWCW13}.

\BB
\section{Concluding Remarks and Open Issues}
\label{sec:Concluding Remarks}
The classical parallel computing systems cannot efficiently process a huge amount of massive data, because of less resiliency to faults and limited scalability of systems. MapReduce, developed by Google in 2004, provides efficient, fault-tolerant, and scalable large-scale data processing at a single site. Hadoop and Spark were not designed for on-site geographically distributed data processing; hence, all the sites send their \emph{raw data} to a single site before a computation proceeds. In this survey, we discussed requirements and challenges in designing geo-distributed data processing using MapReduce and Spark. We also discussed critical limitations of using Hadoop and Spark in geo-distributed data processing. We investigated systems under their advantages and limitations. However, we did not find a system that can provide a solution to all the mentioned challenges in \S\ref{sec:Challenges in the Development of Geo-Distributed Framework}.

\medskip\noindent\textbf{Open issues.} Based on this survey, we identified the following important issues and challenges that require further research:
\begin{itemize}[nolistsep,noitemsep,leftmargin=0.14in]
  \item \emph{Security and privacy}. Most of the frameworks do not deal with security and privacy of data, computation, data transfer, or a deadline-constraint job. Hence, a major challenge for a geo-computation is: how to transfer data and computations to different locations in a secure and privacy-preserving manner, how to trust the requested computations, how to ensure security and privacy within a cluster, and how to meet real-time challenges (recall that we found that G-Hadoop~\cite{g-hadoop}, ViNE~\cite{DBLP:journals/internet/KeaheyTMF09}, and SEMROD~\cite{DBLP:conf/sigmod/OktayMKK15} provide an authentication mechanism, end-to-end data transfer security, and sensitive data security in the hybrid cloud, respectively).

  \item \emph{Fine-grain solutions}. Most of the frameworks do not provide fine-grain solutions to different types of compatibilities. In reality, different clusters have different versions of software, hence, how will be a job executed on different sites having non-identical implementations of MapReduce, operating systems, data storage systems, and security-privacy solutions.

  \item \emph{Global reducer}. There are some solutions (\textit{e}.\textit{g}., G-Hadoop~\cite{g-hadoop} and HMR~\cite{hmr}) that require a global reducer at a pre-defined location. However, the selection of a global reducer has been considered separately while it directly affects the job completion time. Hence, a global reducer may be selected dynamically while respecting several real-time parameters~\cite{DBLP:journals/sigmetrics/GadreRP11}. Though not each site sends its complete datasets, there still exists open questions to deal with, \textit{e}.\textit{g}., should all the DCs send their outputs to a single DC or to multiple DCs that eventually converge, should a DC send its complete output to a single DC or partition its outputs and send them to multiple DCs, and what are the parameters to select a DC to send outputs.

  \item \emph{A wide variety of operations}. The existing work proposes frameworks that allow a limited set of operations. However, it is necessary to find answers to the following question: how to perform many operations like the standard MapReduce on a geographically distributed MapReduce-based framework. Also, we did not find a system that can process secure SQL-queries on geo-distributed data, except in~\cite{sharad-icde-17}, but they focus on the hybrid cloud and store a significant amount of non-sensitive data in the private cloud too.

  \item \emph{Job completion time and inter-DC transfer}. Most reviewed frameworks do not deal with the job completion time. In a geo-distributed computation, the job completion time is affected by distance and the network bandwidth among DCs, the outputs at each DC, and the type of applications. Iridium~\cite{akella-15} and JetStream~\cite{DBLP:conf/nsdi/RabkinASPF14} handle job completion time. However, there is no other framework that jointly optimizes job completion time and inter-DC transfer while regarding variable network bandwidth, which is considered in JetStream~\cite{DBLP:conf/nsdi/RabkinASPF14} and WANalytics~\cite{DBLP:conf/cidr/VulimiriCGKV15}. Thus, there is a need to design a framework that optimizes several real-time parameters and focuses on the job completion time. In addition, the system must dynamically learn and decide whether the phase-to-phase or the end-to-end job completion time is crucial? Answering this question may also require us to find straggling mappers or reducers in the partial or entire computation~\cite{heintz2015end,cardosa2011exploring}.

  \item \emph{Consistency and performance}. A tradeoff is evident between consistency and performance, for example, if a job is distributed over different locations such as in bank transactions. It is required in a geo-distributed computation to have consistent outputs while maximizing the system performance. In order to ensure the output consistency, the distributed components must be in coordination or more appropriately the WAN links must be in coordination. However, achieving coordination is not a trivial task and would certainly incur significant performance overhead in return~\cite{DBLP:conf/sigmod/NawabAA16}.

  \item \emph{Geo-distributed IoT data processing}. We reviewed a sufficient number of stream processing systems. Evidently, there is a huge opportunity in developing real-time stream processing systems for IoT. In an IoT environment, data gathering and real-time data analysis are two prime concerns because of several data outsourcing (sensor) devices, which send small data (\textit{e}.\textit{g}., GPS coordinates) vs large data (\textit{e}.\textit{g}., surveillance videos) possibly at a very high speed. However, the current stream processing systems are not able to handle such a high-velocity data~\cite{karunaratne2016distributed} and require explicit ingestion corresponding to an underlying system~\cite{faisal2017icdcs}. Hence, the existing systems in a geo-distributed IoT system cannot support multiple platforms and underlying databases. In such an environment, it would be interesting to find a way to implement existing popular stream processing systems such as Spark, Flink, and decide how and when to transmit data, which types of algorithms will work regarding small vs large data, how much resources are required at the cloud or edge servers, what would be data filtering criteria, how to maintain privacy of entities, and which DC should be selected for the next level processing.

\item \emph{Geo-distributed machine learning}. Machine learning (ML) provides an ability to analyze and build models from large-scale data. Specifically, ML helps in classification, recommender systems, clustering, frequent itemsets, pattern mining, collaborative filtering, topic models, graph analysis, etc. There are some famous ML systems/libraries, \textit{e}.\textit{g}., Apache Mahout~\cite{apache_mahout}, MLlib~\cite{mllib}, GraphLab~\cite{DBLP:journals/pvldb/LowGKBGH12}, and Google's TensorFlow~\cite{DBLP:conf/osdi/AbadiBCCDDDGIIK16}. However, all these systems deal with only a single DC ML computations. To the best of our knowledge, there are two systems/algorithms, Gaia~\cite{DBLP:conf/nsdi/HsiehHVKGGM17} and~\cite{DBLP:journals/corr/CanoWMCF16}, for performing geo-distributed ML computations. These systems regard variable network bandwidth, and Gaia does not require to change an ML algorithm to be executed over geo-locations. However, we still need to explore a wide variety of geo-distributed ML algorithms in the context of security, privacy, extending MLlib and Mahout to be able to work on geo-distributed settings.
\end{itemize}
In short, we can conclude that geo-distributed big-data processing is highly dependent on the following five factors: task assignment, data locality, data movement, network bandwidth, and security and privacy. However, currently, we are not aware of any system that can jointly optimize all of these factors. In addition, while designing a geo-distributed system, one should memorize the lesson from the experience of Facebook's teams: the system should ``\emph{not just on the ease of writing applications, but also on the ease of testing, debugging, deploying, and finally monitoring hundreds of applications in production}''~\cite{DBLP:conf/sigmod/ChenWIJLSWWWY16}.

\BBB\phantomsection
\section*{Acknowledgements}
The authors of the paper are thankful to the anonymous reviewers for insightful comments. This research was supported by a grant from EMC Corp. We are also thankful to Faisal Nawab for suggesting BigDAWG~\cite{DBLP:conf/hpec/GadepallyCDEHKM16} and Rheem~\cite{DBLP:conf/edbt/AgrawalCEKOPQ0Z16}, Yaron Gonen for discussing early versions of the paper, and Ancuta Iordache for discussing Resilin~\cite{resilin}. The first author's research was partially supported by the Rita Altura Trust Chair in Computer Sciences; the Lynne and William Frankel Center for Computer Science; the grant from the Ministry of Science, Technology and Space, Israel, and the National Science Council (NSC) of Taiwan; the Ministry of Foreign Affairs, Italy; the Ministry of Science, Technology and Space, Infrastructure Research in the Field of Advanced Computing and Cyber Security and the Israel National Cyber Bureau.
\BBB



\begin{IEEEbiography}[{\includegraphics[width=1in,height=1.25in,clip,keepaspectratio]{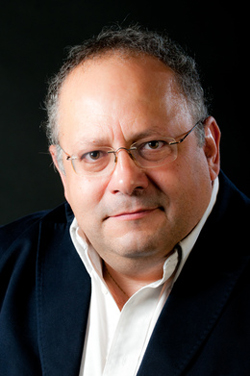}}]{Shlomi Dolev} received his DSc in Computer Science in 1992 from the Technion. He is the founder and the first department head of the
Computer Science Department at Ben-Gurion University, established in 2000. Shlomi is the author of a book entitled Self-Stabilization published by MIT Press in 2000. His publications includes more than three hundred publications. He served in more than a hundred program committees, chairing several including the two leading conferences in distributed computing, DISC 2006, and PODC 2014. Prof. Dolev is the head of the Frankel Center for Computer Science and holds the Ben-Gurion university Rita Altura trust chair in Computer Sciences. From 2011 to 2014, Prof. Dolev served as the Dean of the Natural
Sciences Faculty at Ben-Gurion University of the Negev. From 2010 to 2016, he has served as Head of the Inter University Computation Center of Israel. Shlomi currentlty serves as the steering committee head of the Computer Science discipline of the Israeli ministry of education.
\end{IEEEbiography}
\BBB\BBB\BBB\BBB\BBB

\begin{IEEEbiography}[{\includegraphics[width=1in,height=4.25in,clip,keepaspectratio]{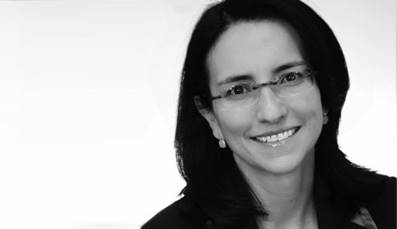}}]{Patricia Florissi} is VP and Global Chief Technology Officer for DellEMC Sales and holds the honorary title of EMC Distinguished Engineer. Patricia is a technology thought leader and innovator, with 20 patents issued and more than 20 patents pending.  Patricia is an active keynote speaker on topics related to big-data, technology trends and innovation. Before joining EMC, Patricia was the Vice President of Advanced Solutions at Smarts in White Plains, New York. Patricia holds a PhD in Computer Science from Columbia University in New York.
\end{IEEEbiography}
\BBB\BBB\BBB\BBB\BBB

\begin{IEEEbiography}[{\includegraphics[width=1in,height=1.25in,clip,keepaspectratio]{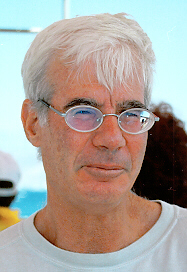}}]{Ehud Gudes} received the BSc and MSc degrees from the Technion and the PhD degree in Computer and Information Science from the Ohio State University in 1976. Following his PhD, he worked both in academia (Pennsylvania State University and Ben-Gurion University (BGU)), where he did research in the areas of database systems and data security, and in industry
(Wang Laboratories, National Semiconductors, Elron, and IBM Research), where he developed query languages, CAD software, and expert systems for planning and scheduling. He is currently a Professor in Computer Science at BGU, and his research interests are knowledge and databases, data security, and data mining, especially, graph mining. He is a member of the
IEEE Computer Society.
\end{IEEEbiography}
\BBB\BBB\BBB\BBB\BBB

\begin{IEEEbiography}[{\includegraphics[width=1in,height=1.25in,clip,keepaspectratio]{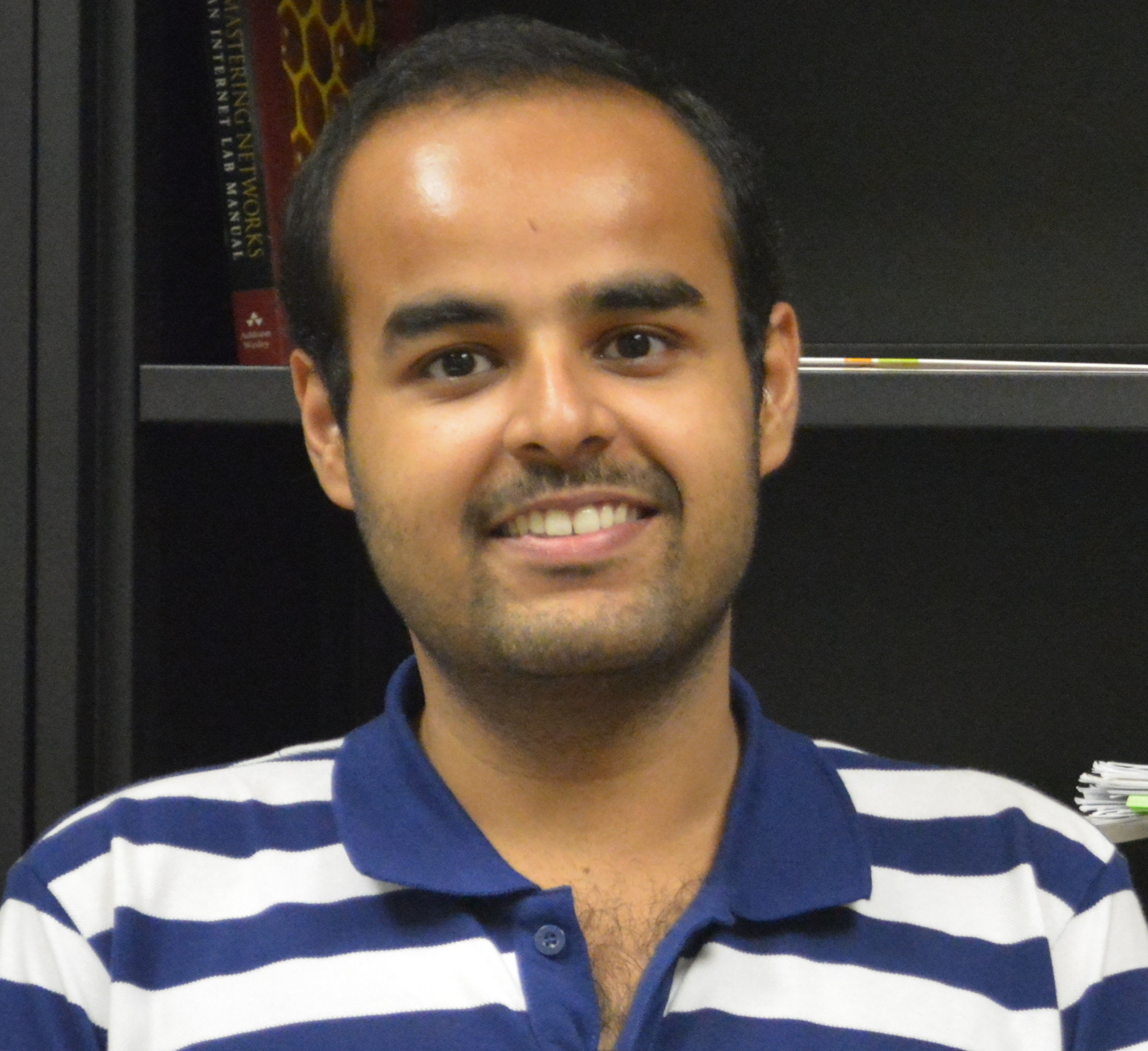}}]{Shantanu Sharma} received his PhD in Computer Science in 2016 from Ben-Gurion University, Israel, and Master of Technology (M.Tech.) degree
in Computer Science from National Institute of Technology, Kurukshetra, India, in 2011. He was awarded a gold medal for the first position in his M.Tech. degree. Currently, he is pursuing his Post Doc at the University of California, Irvine, USA, assisted by Prof. Sharad Mehrotra. His research interests include designing models for MapReduce computations, data security, distributed algorithms, mobile computing, and wireless communication.
\end{IEEEbiography}
\BBB\BBB\BBB\BBB\BBB

\begin{IEEEbiography}[{\includegraphics[width=1in,height=1.25in,clip,keepaspectratio]{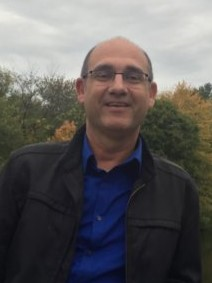}}]{Ido Singer} is a senior software engineering manager in DellEMC Israel. For the past 4 years, he managed the RecoverPoint site in Beer-Sheva, mastering replication data-path algorithms. Ido recently led a big-data innovation collaboration with the Ben-Guryon University. Before joining EMC, Ido worked in the IDF as a system software analyst leading a software transportation algorithm team. Ido holds a PhD in Applied Mathematics from Tel-Aviv University, Israel.
\end{IEEEbiography}

\ifCLASSOPTIONcaptionsoff
  \newpage
\fi

\end{document}